# On the Linear Stability of Magnetized Jets Without Current Sheets –Relativistic Case

By

Jinho Kim[1]★, Dinshaw S. Balsara[1]★, Maxim Lyutikov[2], Sergei S. Komissarov[3]

[1]*Physics Department, College of Science, University of Notre Dame, 225 Nieuwland Science Hall, Notre Dame, IN 46556, USA*

[2]*Department of Physics, Purdue University, 525 Northwestern Avenue, West Lafayette, IN 47907-2036, USA*

[3]*Department of Applied Mathematics, The University of Leeds, Leeds LS2 9GT*

**Abstract**

In our prior papers, we considered the non-relativistic linear stability analysis of magnetized jets that do not have current sheet at the boundary. In this paper, we extend our analysis to relativistic jets. In order to find the unstable modes of current sheet-free, magnetized relativistic jets, we linearize full relativistic magnetohydrodynamics equations and solve them numerically. We find the dispersion relation of the pinch and kink mode instabilities. By comparing the dispersion relations of mildly relativistic jet (Lorentz factor 2) with moderately relativistic jet (Lorentz factor 10), we find that the jet with higher Lorentz factor is significantly more stable in both pinch and kink modes. We show that inclusion of the current sheet-free magnetic field in the jet further enhances the stability. Both pinch and kink mode instabilities become progressively more stable with increasing magnetization. We also show a scaling relation between the maximum temporal growth rate of the unstable mode and the Lorentz factor of the jet. The maximum temporal growth rates of the unstable modes are inversely proportion to the Lorentz factors for most of the modes that we study. However, for the fundamental pinch mode it is inversely proportional to the square of the Lorentz factor. This very beneficial scaling relation holds regardless of the presence of a magnetic field.

★ E-mail: jkim46@nd.edu (JK); dbalsara@nd.edu (DSB)

# 1 Introduction

Relativistic jets are produced in many astrophysical systems such as active galactic nuclei (AGNs), X-ray binaries and gamma-ray bursts (GRBs). The consensus amongst most theorists is that jets form as a result of magnetohydrodynamic (MHD) processes. There are several very prominent theory papers on jet production (Blandford 1976; Lovelace 1976; Blandford & Znajek 1977; Blandford & Payne 1982). In addition to the theoretical works, the numerical studies using general relativistic MHD (GRMHD) simulations support the above scenarios (Koide *et al*. 2000; Nishikawa *et al*. 2005; Hawley & Krolik 2006; McKinney 2006; Komissarov & McKinney 2007; McKinney & Narayan 2007; Komissarov & Barkov 2009; McKinney & Blandford 2009). In both theoretical and numerical papers about jets, the jet is taken to be magnetized. The axial magnetic field is strongly related to the launch of the jet. The hoop stress of the toroidal magnetic field results in collimation of the jets. Observationally, the jets are thought to have very high Lorentz factors. In Lister *et al*. (2009), the extragalactic jets have Lorentz factors up to 50. GRB jets are thought to have even higher Lorentz factors ranging from 100 up to 1000 (Piran 2004; Rykoff *et al*. 2009). Consequently, the jets are strongly magnetized and highly relativistic objects.

One of the puzzling properties of jets is their remarkable stability. The astrophysical jets can propagate the distances of order $10^5 - 10^7$ of their initial radius while terrestrial jets can lose their integrity in the distance of as little as a hundred jet radii. The unstable modes of non-relativistic jets have been thoroughly studied using linear stability analysis (Hardee 1979, 1982; Cohn 1983; Payne & Cohn 1985; Appl & Camenzind 1992; Hardee *et al*. 1992; Appl 1996; Bodo *et al*. 1996; Begelman 1998; Appl *et al*. 2000; Bonanno & Urpin 2011; Kim *et al*. 2015; Bodo *et al*. 2016; Kim *et al*. 2016). In addition, the stability analysis of relativistic jets has been carried out by several authors (Istomin & Pariev 1994, 1996; Lyubarskii 1999; Tomimatsu *et al*. 2001; Narayan *et al*. 2009; Bodo *et al*. 2013). Many of the relativistic stability analyses have been carried out with one or the other simplifying assumption. It is believed that the magnetic field as well as the relativistic effects play significant roles in stabilizing the jets. However, in this paper we demonstrate this very conclusively with stability analysis that is free of simplifying approximations.

In our first paper (Kim *et al*. 2015), we studied the linear stability analysis of the jet with a non-trivial magnetic field proposed by Gourgouliatos *et al*. (2012). The magnetic field does not carry a net-current on the surface of the jet (current sheet-free jet), which has a beneficial effect on jet stability. In Kim *et al*. (2015), we showed that the increasing strength of the magnetic field has a stabilizing effect in both pinch and kink mode instabilities. Our sequel work (Kim *et al*. 2016) considered a velocity shear inside of jets. We studied the stability of magnetized jets with velocity shear and showed that the velocity shear also helps to stabilize the jets. In this paper (our third paper), we study the enhanced stability of relativistic jets. Most of the previous studies of the stability of relativistic jets have considered the force-free magnetic field approximation for the sake of simplicity (Istomin & Pariev 1994; Lyubarskii 1999; Tomimatsu *et al*. 2001; Narayan *et al*. 2009). Or, they assumed a simplified jet with uniform magnetic field so as to make the problem mathematically tractable (Hardee 2007). Bodo *et al.* (2013) solved the full relativistic magnetohydrodynamics equations but they considered a cold gas that does not have thermal pressure. In this paper, we solve the full linearized relativistic magnetohydrodynamics equations with non-trivial magnetic field, which can only be done numerically. This is accomplished using the numerical technique in Kim *et al*. (2015).

The remainder of the paper is divided as follows. In Section 2 we derive the governing equations for linear stability analysis of jets with non-trivial magnetic field. In Section 3 we describe the baseline model of the unperturbed jet in this paper. In Section 4 we compare the linear stability of jets that have mild Lorentz factor with jets that have moderate Lorentz factor. In Section 5 we present the role of the magnetization in the stability of the jets. In Section 6 we compare the linear stability of relativistic jets with increasing magnetization. Section 7 presents discussion and conclusions.

## 2 Linearized Equations and Their Solution

In this paper we consider ideal relativistic flows. We assume an isentropic equation of state. In other words, we use a polytropic equation of state ($P \sim \rho^\Gamma$) instead of solving the linearized energy equation for the sake of simplicity. Then, the conservative form of relativistic MHD equations in cylindrical coordinate in flat spacetime are expressed as (Anton *et al.* 2006)

$$\frac{\partial \rho\gamma}{\partial t}+\frac{1}{r}\frac{\partial}{\partial r}(r\rho\gamma v_r)+\frac{1}{r}\frac{\partial}{\partial \phi}(\rho\gamma v_\phi)+\frac{\partial}{\partial z}(\rho\gamma v_z)=0, \tag{1}$$

$$\frac{\partial S_r}{\partial t}+\frac{1}{r}\frac{\partial}{\partial r}\left[r\left(S_r v_r+\Pi-b_r B_r/\gamma\right)\right]+\frac{1}{r}\frac{\partial}{\partial \phi}\left(S_r v_\phi-b_r B_\phi/\gamma\right)+\frac{\partial}{\partial z}\left(S_r v_z-b_r B_z/\gamma\right)$$
$$=\frac{1}{r}\left[\left(\rho h+b^2\right)\gamma^2 v_\phi^2+\Pi-b_\phi^2\right] \tag{2}$$

$$\frac{\partial S_\phi}{\partial t}+\frac{1}{r^2}\frac{\partial}{\partial r}\left[r^2\left(S_\phi v_r-b_\phi B_r/\gamma\right)\right]+\frac{1}{r}\frac{\partial}{\partial \phi}\left(S_\phi v_\phi+\Pi-b_\phi B_\phi/\gamma\right)+\frac{\partial}{\partial z}\left(S_\phi v_z-b_\phi B_z/\gamma\right)=0, \tag{3}$$

$$\frac{\partial S_z}{\partial t}+\frac{1}{r}\frac{\partial}{\partial r}\left[r\left(S_z v_r-b_z B_r/\gamma\right)\right]+\frac{1}{r}\frac{\partial}{\partial \phi}\left(S_z v_\phi-b_z B_\phi/\gamma\right)+\frac{\partial}{\partial z}\left(S_z v_z+\Pi-b_z B_z/\gamma\right)=0, \tag{4}$$

$$\frac{\partial B_r}{\partial t}+\frac{1}{r}\frac{\partial}{\partial \phi}\left(B_r v_\phi-B_\phi v_r\right)+\frac{\partial}{\partial z}\left(B_r v_z-B_z v_r\right)=0, \tag{5}$$

$$\frac{\partial B_\phi}{\partial t}+\frac{\partial}{\partial r}\left(B_\phi v_r-B_r v_\phi\right)+\frac{\partial}{\partial z}\left(B_\phi v_z-B_z v_\phi\right)=0, \tag{6}$$

$$\frac{1}{r}\frac{\partial (rB_r)}{\partial r}+\frac{1}{r}\frac{\partial B_\phi}{\partial \phi}+\frac{\partial B_z}{\partial z}=0. \tag{7}$$

Here, $\gamma=1/\sqrt{1-\mathbf{v}^2}$ is Lorentz factor, $S_i=\left(\rho h\gamma^2+B^2\right)v_i-(\mathbf{B}\cdot\mathbf{v})B_i$ is momentum density and $b_i=B_i/\gamma+\gamma(\mathbf{B}\cdot\mathbf{v})v_i$ gives the spatial components of the magnetic four vector. The total pressure (gas+magnetic) is defined as $\Pi=P+b^2/2$, where the amplitude of the magnetic four vector is $b^2=B^2/\gamma^2+(\mathbf{B}\cdot\mathbf{v})^2$. Note that the $(4\pi)^{1/2}$ factor is absorbed in the definition of magnetic field. We take advantage of divergence free nature of magnetic field to eliminate the $B_z$ evolution equation. Thus we use the divergence-free constraint of the magnetic field (Eq. 7). In cylindrical coordinates, the unperturbed jet solution can be described by the functions which are only dependent on the radial coordinate- $\rho_0(r)$, $v_{z0}(r)$, $P_0(r)$, $B_{z0}(r)$, $B_{\phi 0}(r)$ are the rest mass density, axial velocity, pressure, axial and azimuthal magnetic field respectively. The unperturbed radial and rotational velocity ($v_{r0},v_{\phi 0}$) and the radial magnetic field ($B_{r0}$) are zero in our model of the jet.

In this paper, we use non-trivial structures for the magnetic field which are solutions of the steady-state MHD equations (also known as the Grad-Shafranov equation). In these solutions, the gradient of total pressure force in the unperturbed jet is exactly balanced by the magnetic tension force of the toroidal magnetic field. In this paper, we also assume that the unperturbed ambient medium is non-magnetized, uniform and stationary. The enthalpy and Lorentz factor of unperturbed state can be derived from the above variables. i.e., $h_0(r) = 1 + \Gamma P_0(r) / \left( (\Gamma-1) \rho_0(r) \right)$ and $\gamma_0(r) = 1/\sqrt{1 - v_{z0}^2(r)}$.

As in our previous study, the perturbation of the variables has a form of $\delta f(t, r, \phi, z) = \delta f(r) \exp\left[ i(\omega t - m\phi - kz) \right]$. We only use $m=0$ for pinch mode and $m=1$ for kink mode. We do not examine models with $m>1$ in this paper because prior work has shown that the most unstable modes are indeed the pinch and kink modes. We only consider the complex number solution of $\omega$ as a function of real $k$. If $\omega$ has a negative imaginary part, it means that the solution has an unstable and growing mode. The spatial growth rates, which is the ratio between the temporal grow rate and the group velocity, sometimes is used in the jet stability analysis (Drazin & Reid 1981). We, however, do not take the spatial growth rate into account in this paper. We make the further definition $\tilde{\omega}(r) \equiv \omega - v_{z0}(r) k$ and $k_B(r) = \dfrac{m}{r} B_{\phi 0}(r) + k B_{z0}(r)$. $\tilde{\omega}$ is usually called the Doppler shifted frequency. The MHD equations, the polytropic equation of state, and the divergence free condition ($\nabla \cdot \mathbf{B} = 0$) give us the following linearized equations.

The continuity equation gives

$$i\tilde{\omega}(r) \frac{\delta \rho}{\rho_0(r)} + \frac{1}{r \rho_0(r) \gamma_0(r)} \frac{d}{dr}\left( r \rho_0(r) \gamma_0(r) \delta v_r \right) - i\frac{m}{r} \delta v_\phi + i\tilde{\omega}(r) \gamma_0^2(r) v_{z0}(r) \delta v_z - ik\, \delta v_z = 0 \quad . \tag{8}$$

The $r$-momentum equation gives

$$i\tilde{\omega}\left(\rho_0(r)h_0(r)\gamma_0^2(r)+B_{\phi 0}^2(r)\right)\delta v_r + i\left(\omega B_{z0}(r)+\frac{m}{r}B_{\phi 0}(r)v_{z0}(r)\right)\left(B_{z0}(r)\delta v_r - v_{z0}(r)\delta B_r\right)$$
$$+\frac{2}{r}B_{\phi 0}(r)\left[\left(1-v_{z0}^2(r)\right)\delta B_\phi - B_{\phi 0}(r)v_{z0}(r)\delta v_z + B_{z0}(r)v_{z0}(r)\delta v_\phi\right]+ik_B\delta B_r+\frac{d\delta\Pi}{dr}=0 \quad . \tag{9}$$

The angular momentum conservation equation gives

$$i\tilde{\omega}\rho_0(r)h_0(r)\gamma_0^2(r)\delta v_\phi$$
$$+i\omega\left(B_{z0}^2(r)\delta v_\phi - B_{\phi 0}(r)B_{z0}(r)\delta v_z - B_{z0}(r)v_{z0}(r)\delta B_\phi - B_{\phi 0}(r)v_{z0}(r)\delta B_z\right)$$
$$-\frac{2im}{r}\left[B_{\phi 0}^2(r)v_{z0}(r)\delta v_z - B_{\phi 0}(r)B_{z0}(r)v_{z0}(r)\delta v_\phi - \left(1-v_{z0}^2(r)\right)B_{\phi 0}(r)\delta B_\phi\right] \quad . \tag{10}$$
$$-\frac{im}{r}\delta\Pi + ik\left(B_{\phi 0}(r)\delta B_z + B_{z0}(r)\delta B_\phi\right)$$
$$-\frac{1}{r^2}\frac{d}{dr}\left[r^2\left(B_{\phi 0}(r)B_{z0}(r)v_{z0}(r)\delta v_r + \left(1-v_{z0}^2(r)\right)B_{\phi 0}(r)\delta B_r\right)\right]=0$$

The $z$-momentum equation gives

$$i\tilde{\omega}\left[\begin{array}{l}\rho_0(r)h_0(r)\gamma_0^2(r)\left(1+2\gamma_0^2(r)v_{z0}^2(r)\right)\delta v_z + B_{\phi 0}^2(r)\delta v_z \\ +\gamma_0^2(r)v_{z0}(r)\left(\delta\rho+\frac{\Gamma}{\Gamma-1}\delta P\right)+2B_{\phi 0}(r)v_{z0}(r)\delta B_\phi - B_{\phi 0}(r)B_{z0}(r)\delta v_\phi\end{array}\right]$$
$$+\frac{1}{r}\frac{d}{dr}\left[r\left(\left(\rho_0(r)h_0(r)\gamma_0^2(r)+B_{\phi 0}^2(r)\right)v_{z0}(r)\delta v_r - B_{z0}(r)\delta B_r\right)\right] \quad . \tag{11}$$
$$-i\rho_0(r)h_0(r)\gamma_0^2(r)v_{z0}(r)\left(\frac{m}{r}\delta v_\phi + k\delta v_z\right)+k_B\delta B_z$$
$$+iB_{z0}(r)\left(\frac{m}{r}\delta B_\phi + k\delta B_z\right)+ik\left[B_{\phi 0}(r)B_{z0}(r)v_{z0}(r)\delta v_\phi - \delta\Pi\right]=0$$

The $r$-magnetic field equation gives

$$i\tilde{\omega}\delta B_r + ik_B\delta v_r = 0 \quad . \tag{12}$$

The $\phi$-magnetic field equation gives

$$i\tilde{\omega}\delta B_\phi + \frac{d}{dr}\left(B_{\phi 0}(r)\delta v_r\right)-ik\left(B_{\phi 0}(r)\delta v_z - B_{z0}(r)\delta v_\phi\right)=0 \quad . \tag{13}$$

The divergence free constraint of magnetic field gives

$$\frac{1}{r}\frac{d}{dr}(r\delta B_r) - i\frac{m}{r}\delta B_\phi - ik\delta B_z = 0. \tag{14}$$

The equation of state provides

$$\frac{\delta P}{P_0(r)} = \Gamma \frac{\delta \rho}{\rho_0(r)}. \tag{15}$$

Finally, the linearization of the definition of the total magnetic field gives

$$\delta\Pi = \delta P + B_{\phi 0}(r)\delta B_\phi + B_{z0}(r)\delta B_z - B_{\phi 0}(r)v_{z0}(r)\big(v_{z0}(r)\delta B_\phi + B_{\phi 0}(r)\delta v_z - B_{z0}(r)\delta v_\phi\big). \tag{16}$$

Note that the subscript 0 denotes the unperturbed variables.

The linearized equations Eqs. (8)- (16) consist of six differential equations and three algebraic equations. To eliminate the differential equation as much as possible, we first obtain the expressions of $d\delta v_r/dr$ and $d\delta B_r/dr$ from Eq. (8) and Eq. (13). Then, $d\delta v_r/dr$ and $d\delta B_r/dr$ in Eqs. (10)- (13) are replaced by the above expressions. As a result, we finally have two differential equations and seven algebraic equations, which is exactly same form of equations as those of the non-relativistic jet stability analysis (Kim *et al.* 2015). For that reason, we do not repeat our description of the numerical method used in this paper. However, the matrix components for solving the algebraic equations and the asymptotic behaviors of $\delta v_r$ and $\delta\Pi$ near the jet axis ($r=0$) should be rewritten in the relativistic form. They are presented in the Appendices A and B.

Before going into the next section, we need to briefly describe the boundary condition of our system of equation. Just like non-relativistic jet stability analysis (Kim *et al.* 2015), the solution of ambient medium has the analytic form of $\delta P = K_{|m|}(\kappa r)$, where $\kappa^2 = k^2 - \omega^2/c_s^2$. Notice that the sound speed here is the relativistic sound speed ($c_s^2 = \Gamma P_a/(\rho_a h_a)$). Then the continuity across the jet boundary of the total pressure ($\delta\Pi$) and radial displacement ($\delta v_r/(i\tilde{\omega})$) gives the dispersion relation of the jet stability:

$$\frac{i\tilde{\omega}\delta\Pi(r=1)}{\delta v_r(r=1)} = \frac{\rho_a h_a \omega^2 K_{|m|}(\kappa)}{\kappa K'_{|m|}(\kappa)} \ . \tag{17}$$

The non-magnetized jet model has uniform density and pressure. In this case, we can easily obtain the analytic solution inside of the jet using the (modified) Bessel function. All the details will appear in the Appendix C. We also present the marginal stability points where the reflections modes become unstable in Appendix C.

### 3 Baseline Models for Unperturbed Jets

All the magnetized jet models that we consider in this paper do not carry a net electric current inside the jet. Gourgouliatos *et al*. (2012) present a special configuration of magnetic field with this very favorable property. In this model, both axial and toroidal magnetic field vanish at the surface of the jet. As a result, the jet model does not have a current sheet at the boundary. Gourgouliatos *et al*. (2012) propose the axial and toroidal magnetic field as a function of radius. Then the gas pressure is provided by the Grad-Shafranov equation for the proposed magnetic field structure. This is shown Eqs. (24)-(26) in Gourgouliatos *et al*. (2012). Fig. 1(a) shows the poloidal and toroidal magnetic field structure that is used in this paper. The corresponding gas pressure is shown in Fig. 1(b). Notice that the toroidal magnetic field ($B_\phi$) is proportional to the Lorentz factor of the jet since Gourgouliatos *et al*. (2012) assumed that it is measured by the Eulerian observer.

Although we do not take resistive effects into account in this paper, this current-sheet-free configuration of magnetic field has several merits. Since there is no resistive type of instability at the surface of the jet, the jet becomes more stable. Moreover, the magnetic field that they proposed does not have surface force at the boundary of the jet. For this reason, the current sheet free configuration of magnetic field is very desirable and suitable for the non-linear numerical simulation which will be part of our subsequent work.

In Kim *et al.* (2015, 2016), the jet models are parameterized by the ratio of the jet and ambient densities $\eta=\rho_j/\rho_a$, the jet Mach number $M=v_j/c_s$, where $c_s$ is the sound speed, and the plasma beta $\beta=P_j/(B^2_j/2)$. Here, the subscripts *j* and *a* denote the jet and ambient medium, respectively. All the jet's physical quantities with subscript *j* except the speed of jet ($v_j$) are measured in the comoving frame of the jet. We keep the jet parameters entirely consistent with the previous paper. For the non-uniform (magnetized) jet, the values of above parameters are different in different locations. We use the parameter values that are measured at the axis of the jet. Notice that in the rest frame of the jet, the highly relativistic jet ($\gamma_0 \gg 1$) has very strong toroidal magnetic field which is proportional to the Lorentz factor (see Eq. (24) in Gourgouliatos *et al.* (2012)). However, the magnetic pressure of our jet model is expressed as $P_{mag} = \left( B_{\phi 0}^2 / \gamma_0^2 + B_{z0}^2 \right)/2$. This expression of the magnetic pressure becomes equivalent with the magnetic pressure in the comoving frame.

Here are the actual values of the parameters that are used throughout the paper. i) The polytropic exponent, $\Gamma$ is 5/3. ii) The jet and ambient density ratio, $\eta$, is 0.1. iii) The Mach number, $M$, is 4. iii) The unperturbed jet velocity has a top hat profile. The jet does not have velocity shear inside. iv) We use several different plasma beta, $\beta$ values. They are $\infty$, 1, 1/2 and 1/4 from the non-magnetized jet to the very strongly magnetized jet. In addition to the plasma beta, the magnetization, $\sigma = B_j^2/(\rho_j h_j)$ and the ratio the Alfven speed to the sound speed, $M_a = c_a/c_s$ are also widely used to parametrize the strength of the magnetic field in the jets. Here, $c_a$ is the Alfven speed and defined as $c_a = \sqrt{B_j^2 / (\rho_j h_j + B_j^2)} = \sqrt{\sigma/(1+\sigma)}$. Their values of the $M=4$ jets with various plasma beta are provided in Table 1.

.

|  | $\gamma_0=2$ |  | $\gamma_0=10$ |  |
| --- | --- | --- | --- | --- |
| $\beta$ | $\sigma(\times 10^{-1})$ | $M_a$ | $\sigma(\times 10^{-1})$ | $M_a$ |
| $\infty$ | 0 | 0 | 0 | 0 |
| 1 | 0.563 | 1.07 | 0.743 | 1.06 |
| 1/2 | 1.13 | 1.47 | 1.49 | 1.45 |
| 1/4 | 2.25 | 1.98 | 2.97 | 1.92 |

Table 1. The $\sigma$ and $M_a$ values of the $M=4$ jets with various plasma $\beta$.

When comparing relativistic with non-relativistic jets, the definition of Mach number may become confusing. Many researches in the past just used the ratio $v_j/c_s$ to define the Mach number in the relativistic regime. Although the traditional choice of Mach number is suitable for parametrization, this definition does not relate to the wave propagation in the same way in the non-relativistic and relativistic regimes. If we want to preserve this relation (e.g. the relation between the Mach number and the Mach angle) then the generalized definition of the Mach number should be (proper Mach number, Konigl 1980; Komissarov & Falle 1998)

$$M_{rel} = \frac{v_j / \sqrt{1-v_j^2}}{c_s / \sqrt{1-c_s^2}}. \tag{18}$$

With this definition of $M_{rel}$, a higher Lorentz factor implies a higher Mach number, even if the traditional Mach number, $M = v_j / c_s$, remains fixed. The $M_{rel}$ values of $M=4$ jets when the jets' Lorentz factors are 2, 10, 100 and 1000 are given in Table 2.

| $\gamma_j$ | 2 | 10 | 100 | 1000 |
|---|---|---|---|---|
| $M_{rel}$ | 7.810 | 38.74 | 387.3 | 3873 |

Table 2. The $M_{rel}$ values defined in Eq. (18) of $M=4$ jet with various jet Lorentz factor.

Any jet always has some amount of instability. It is very intuitively helpful to understand the jet stability if we make a stability criterion for the jet. Kim *et al.* (2016) consider such a criterion of the jets' instability. This criterion is obtained by comparing the e-folding time due to the instabilities of the jets with the time taken by the jet to propagate hundreds of jets' radii. The e-folding time of the jet is the inverse of the temporal growth rate, $\tau = 1/|\omega_I|$. And the time taken by the jet's unstable modes which propagate with their group velocities ($v_g$) to reach $\chi$ jet radii is given by $T = \chi r_j / v_g$. Accordingly, we say that the instability will not destabilize the jet if $\tau \geq T$. The equivalent expression is

$$\frac{\omega_I r_j}{c_s} \leq \frac{v_g}{\chi c_s}. \tag{19}$$

For this paper, we use $\chi = 400$. In other words, a jet might be "quite stable" if the perturbation of jet can propagate 400 or more jet radii before one e-folding time. All the figures of the dispersion relations after this point have dotted lines which indicate the thresholds of the stability criterion of each unstable mode. To evaluate group velocity ($v_g = d\omega_R/dk$), we numerically differentiated the real part of $\omega$ with respect to the wave number $k$.

## 4 Stability of Jets with Different Lorentz Factors

### 4.1 Improving Stability with Increasing Lorentz Factor

In order to describe how the stability of the jet is affected by its relativistic motion, we carry out a linear stability analysis of jets with mild ($\gamma_0=2$) and moderate ($\gamma_0=10$) Lorentz factors. The corresponding unperturbed jet velocity is $v_{z0}=0.866$ and $v_{z0}=0.995$, respectively. The higher speed jet has a larger central sound speed, so as to keep the Mach number constant at 4.

Fig. 2 shows the dispersion relation of the pinch mode (m=0) instabilities for the non-magnetized relativistic jets. The solid and dashed lines represent the real and imaginary part of $\omega$, respectively. We show only the fundamental and the first reflection modes in Fig 2. Fig 2(a) shows the instabilities of the mild Lorentz factor ($\gamma_0=2$) jet while Fig. 2(b) shows the instabilities of the moderate Lorentz factor ($\gamma_0=10$) jet. Let's compare Fig. 2(a) with Fig. 2(b) to see the effect of increasing Lorentz factor on the stability of jets. The moderate relativistic jet ($\gamma_0=10$) has a significantly lower temporal growth rate for the fundamental pinch mode compared to the mild relativistic ($\gamma_0=2$) jet. This difference is more than one order of magnitude at all the wavelengths!

The comparison of the first reflection pinch mode in Fig. 2 shows that the reflection mode is also stabilized by the relativistic motion. The first reflection mode in Fig. 2(b) is considerably more stable than the first reflection mode in Fig. 2(a) for several different wavelengths. Although the first reflection mode shows a similar trend with the fundamental pinch mode instability, there is one distinguishable property in the reflection mode instability. The first reflection mode of the moderate relativistic ($\gamma_0=10$) jet starts to be destabilized at a longer wavelength compared the mild relativistic ($\gamma_0=2$) case. I.e., the marginal stability point of the moderate relativistic jet is located at the longer wavelength than that of the mild relativistic jet. This fact will be discussed later in this section.

Fig. 3 is analogous to Fig. 2 but for the *m*=1 kink modes. The kink mode instabilities are thought to very easily destabilize the jet and destroy the jet structure. Accordingly, they are usually regarded as the most dangerous instability mode of the jet. (To make matters worse, the destabilization usually occurs even for the magnetized jets due to the current driven instabilities which are dominant in the kink modes.) Many previous studies have shown special concern for the kink mode instabilities (Appl & Camenzind 1992; Istomin & Pariev 1994, 1996; Begelman 1998; Lyubarskii 1999; Tomimatsu *et al.* 2001; Narayan *et al.* 2009; Bodo *et al.* 2013, 2016). This fact is easily observed in Fig 3. Both the fundamental and the first reflection modes are above the stability threshold even for the jet with higher Lorentz factor ($\gamma_0$=10). A comparison Fig. 3 (a) with (b) shows that the relativistic effects enhance the stability of the kink fundamental mode at long as well as short wavelengths. However, the stabilizing effect is not significant at long wavelengths ($kr_j$<$10^{-1}$). On the other hand, the growth rate at short wavelength for the higher Lorentz factor jet ($\gamma_0$=10) is over one order of magnitude smaller than the growth rate of the $\gamma_0$=2 jet.

The first reflection kink instability in Fig. 3 shows almost similar behavior to the pinch instability in Fig. 2 and the difference in the temporal growth rate for the $\gamma_0$=10 jet is also around one order of magnitude compared to the $\gamma_0$=2 jet. The locations of the marginal stability points of both mild and moderate relativistic jets also show the same trend with the first reflection pinch instabilities.

In Figs. 2-3, there are mainly two interesting features in both pinch and kink mode instabilities. One is that the jet with a higher Lorentz factor (moderate relativistic jet) is much more stable in both fundamental and reflection modes. This fact is exactly consistent with the previous studies of relativistic unmagnetized jets (Ferrari, Trussoni & Zaninetti 1978; Hardee 1979). It can be explained by the following fact. Since the higher Lorentz factor jet has effectively higher inertia, (i.e., inertia goes as Lorentz factor squared) it is much less influenced by the Kelvin-Helmholtz instability occurring at the vortex sheet (jet-ambient boundary). In other words, the energy loss of the higher Lorentz factor jet from the vortex sheet compared with its kinetic energy is significantly smaller. The same trend also applies to the reflection modes--- the higher Lorentz factor jets have lower growth rates.

The other trend that we observe in Figs. 2 and 3 is that the reflection modes start from much lower *k* value for the moderate Lorentz factor jets ($\gamma_0$=10). This is somewhat contradictory to our intuition that the relativistic jet is more stable. It can be explained by the analytic dispersion relation of the non-magnetized relativistic jet, which appears in the Appendix C. Appendix C is a relativistic extension of the work by Cohn (1983) in the non-relativistic jet stability analysis. According to Cohn (1983), there are infinite number of marginal stability points where both real and imaginary part of $\omega$ become zero. Eq. (22) in Cohn (1983) gives a precise value for the onset of the reflection mode instability. In Appendix C, we have obtained an analogous expression for the relativistic jet (Eq. (C2)). In our analytic dispersion relation for the non-magnetized relativistic jet, the marginal stability point wave number $k_n$ for the n-th reflection mode is inversely proportional to the Lorentz factor $\gamma_0$. We have marked a green arrow at the marginal stability point of the first reflection mode that we found from the analytic solution in Figs 2 and 3.

**4.2 Comparison of Relativistic Jets with Non-Relativistic Jets**

Let us also compare our mildly relativistic ($\gamma_0$=2) results with non-relativistic results. To do that, please compare Fig. 2 with Fig. 4(a). Fig. 4(a) in this paper is actually borrowed from Kim *et al*. (2015), where it was labeled as Fig 2(a). Fig 4(a) shows a dispersion relation of the non-relativistic and non-magnetized jet with $\eta$=0.1 and $M$=4. It can, therefore, be directly compared to Fig. 2. We see that the fundamental pinch mode in Fig 2(a) for the mildly relativistic ($\gamma_0$=2) case is slightly more stable than the fundamental pinch mode in Fig. 4(a). Similarly, the first reflection pinch mode is also slightly more stable for the mildly relativistic ($\gamma_0$=2) jet. The fundamental pinch mode in Fig 2(b) for the moderately relativistic ($\gamma_0$=10) case is dramatically stabilized and there is no range of wavelength that is above the stability criterion. Furthermore, we only see a narrow range which is above the stability criterion for the first reflection mode.

Again, let us also compare our relativistic results with non-relativistic results for the kink (m=1) modes. To do that, please compare Fig. 3 with Fig. 4(b). Fig. 4(b) in this paper was labeled as Fig. 2(b) in Kim *et al*. (2015). The fundamental kink mode in Fig 3(a) for the mildly relativistic ($\gamma_0$=2) jet also shows slightly better stability at short wavelength compared to the fundamental kink mode in Fig. 4(a). Similarly, the first reflection kink mode is also slightly more stable at short wavelength for the mildly relativistic jet ($\gamma_0$=2). For the moderately relativistic ($\gamma_0$=10) case in Fig. 3(b), we can observe that the fundamental as well as the reflection kink mode is substantially stabilized at short wavelength.

In Fig. 4, the marginal stability point for the non-relativistic work is located at even higher

$k$ value compared to the mildly relativistic case. It is also explained by the non-relativistic limit ($\gamma_0 \to 1$) of the analytic dispersion relation in Eq. (C2).

**4.3 Do Higher Lorentz Factors Provide Substantially Greater Stability?**

From the comparison study of the mildly and moderately relativistic jets, we observe that higher Lorentz factor jets have better stability compared to the lower Lorentz factor jets. In this subsection, we will show how the trend behaves with increasing Lorentz factor. To do that, we pick up the maximum growth rate (the growth rate at resonance) of the fundamental as well as the first reflection modes for the jets with various Lorentz factors from 1 all the way up to 5000. Fig. 5 shows the maximum growth rate as a function of the Lorentz factor on a logarithmic scale. Analogous to the previous figures, the red and blue lines represent the fundamental and the first reflection mode. Fig. 5(a) shows the pinch mode ($m=0$) instabilities. For the fundamental pinch mode, there is not an apparent resonant point. We, therefore, use the growth rate at $kr_j=1$ where the growth rate has almost constant and maximum value. In Fig. 5(a), we can clearly see the trend that the jet with higher Lorentz factor has better stability continues to the extremely high Lorentz factors. We also found the empirical scaling law of the stability for the highly relativistic jets. For the fundamental pinch mode, the temporal growth rate is inversely proportional to the square of the Lorentz factor ($\omega_I \sim 1/\gamma_0^2$). However, the growth rate of the first reflection pinch mode is just proportional to $1/\gamma_0$. Hence, the reflection modes of the highly relativistic jets are dominant for the pinch mode instability.

Fig. 5(b) shows the fundamental and the first reflection kink mode ($m=1$) instability of the jets with various Lorentz factors. In Fig. 5(b), we again can see the trend that the jet with higher Lorentz factor has better stability. However, the empirical scaling law of the stability for the highly relativistic jets is different. For the kink instability, both fundamental and reflection modes have the same scaling law. The growth rate is inversely proportional to the Lorentz factor ($\sim 1/\gamma_0$). As a result, the fundamental mode as well as the reflection modes of the highly relativistic jets are dominant in the kink mode instability.

Hardee (2007) have shown the analytic dispersion relation of the jet that has a uniform density, pressure and poloidal magnetic field profile. In Hardee (2007), the dependences of the growth rate on the Lorentz factor are shown for the very limited cases such as at a long wavelength limit or at a resonance point. The Eqs. (5), (8) and (B26) in Hardee (2007) is completely consistent with our scaling relation. Fig. 9 in Bodo *et al.* (2013) also have shown that the relativistic motion of the jet significantly stabilizes the jet. Although Bodo *et al.* (2013) have shown this property up to Lorentz factor of 10, our result shows a good agreement with their work while vastly extending the range of Lorentz factors that are considered.

### 4.4 Implications for the Onset of Turbulence

The development of turbulence in astrophysics is a completely non-linear phenomenon. Although it is not possible to discuss about turbulence within the context of a linear stability analysis, we can at least think about the onset of the turbulence of the relativistic jet in this paper. If the jet starts off with a Laminar flow, then the only source of turbulence in the jet will be due to the short wavelength instabilities that propagate into the jet from its boundary. We have seen that these short wavelength instabilities are suppressed for the strongly relativistic jet in both the pinch as well as the kink modes. Based on these facts, we suggest that the highly relativistic jet might be less susceptible to develop small scale turbulence.

## 5 Effect of Magnetization

### 5.1 Improving Stability with Increasing Lorentz Factor

Let us focus on the stability of the magnetized jet in this section. The unperturbed jet from the previous section is now strongly magnetized ($\beta$=1/2). In other words, the magnetic pressure is twice as strong as the gas pressure. Again, in this section, we mainly make comparisons of the stability of jets having different Lorentz factors.

Fig. 6 shows the dispersion relation of the pinch mode ($m$=0) instabilities for the strongly magnetized relativistic jets. Fig 6(a) shows the instabilities of the mild Lorentz factor ($\gamma_0$=2) jet while Fig. 6(b) shows the instabilities of the moderate Lorentz factor ($\gamma_0$=10) jet. By comparing Fig. 6 with Fig. 2, we see that the magnetic field dramatically stabilizes the fundamental mode of the pinch instability at short as well as long wavelengths. The stabilizing effect is much more evident at short wavelength. There is no unstable fundamental mode observed at short wavelength ($kr_j$>0.3~0.4) for the jets with both mild and moderate Lorentz factors. In our non-relativistic jet stability study (Kim *et al.* 2015), we claimed that the strong magnetic field without current sheet can form ranges of stability where the unstable modes are absent. The same trend appears in this relativistic work. This provide a useful concordance between non-relativistic and relativistic jets with magnetization.

The first reflection modes in Fig 6 are not much influenced by the magnetic field. Their overall growth rate of the first reflection mode is not much changed compared to the non-magnetized jet case in Fig. 2. The marginal stability points also remain almost unchanged.

Fig. 7 is analogous to Fig. 6 but for the kink mode ($m$=1) instabilities. The strong magnetic field leads to substantial improvement of the stability at short wavelength for the fundamental kink mode. Just as in Fig. 6, there is a range of wave numbers with $kr_j$>1~2 for which the fundamental kink instability mode does not exist. However, a significant stabilizing effect is not observed in the fundamental kink mode instability at long wavelength. Please compare Fig 3 with Fig 7.

The strong magnetic fields also stabilize the short wavelengths of the first reflection kink mode, which is shown in the mild Lorentz factor jet ($\gamma_0$=2); please see Fig. 7(a). Like the first reflection pinch instability, the marginal stability points also remain almost unchanged. Further investigation of stabilizing effects due to increasing magnetic field strength will be discussed in section 6.

Let us compare the stability of relativistic magnetized jets with their non-relativistic counterparts. Fig. 8 shows the non-relativistic stability analysis of the strongly magnetized jet. Fig. 8 of this paper is borrowed from Kim *et al*. (2015), where it was labeled as Fig 5. We can directly compare Fig. 6 with Fig 8(a). We see that the fundamental pinch mode in Fig 6(a) for the mildly relativistic ($\gamma_0$=2) case is slightly more stable than the fundamental pinch mode in Fig. 8(a). The growth rate of the fundamental pinch mode even for the mildly relativistic jet is now below the stability criterion. For the first reflection pinch instability mode, the growth rate at resonance (the wave number where its growth rate is a maximum) has a slightly lower value. However, the difference is not significant.

Similarly, let us also compare our relativistic results with non-relativistic results for the kink (*m*=1) modes. Please compare Fig. 7 with Fig. 8(b). Actually there are only minor improvements in the stability for the fundamental kink mode of the mildly relativistic jet. For the first reflection mode, there is only a very tiny range of unstable wavelengths in the non-relativistic jet. However, the range is relatively wider in the relativistic jet.

### 5.2 Scaling Property of the Strongly Magnetized Jets

Analogously to section 4.3, we study the stability of magnetized jets with increasing Lorentz factors. In this subsection, we use the strongly magnetized jets ($\beta$=1/2) to verify that there still exists a scaling law even in the magnetized case.

Fig. 9 is analogous to Fig. 5 but for the strongly magnetized jet case. Fig. 9(a) shows the pinch mode (*m*=0) instabilities. In Fig. 9(a), we can see the same trend which is shown in Fig 5(a). The temporal growth rate of the fundamental pinch instability mode is inversely proportional to the square of the Lorentz factor ($\omega_I \sim 1/\gamma_0^2$). The growth rate of the first reflection pinch instability mode, however, is just proportional to $1/\gamma_0$.

Fig. 9(b) shows the fundamental and the first reflection kink mode (*m*=1) instability of the jets with various Lorentz factors. Just like the non-magnetized jet case, the empirical scaling law of the stability for the highly relativistic jets follows the same trend in Fig. 5(b). Both fundamental and reflection modes have the same scaling law and the temporal growth rate is inversely proportional to the Lorentz factor ($\sim 1/\gamma_0$).

In this section, we can observe that regardless the presence of the magnetic field in the jet, the highly relativistic jets are very stable. The temporal growth rates of the instability modes are either inversely proportion to the Lorentz factors or inversely proportional to the square of the Lorentz factor.

### 6 Stability of Jets with Different Magnetizations

Now let us focus on the stability analysis of relativistic jets with various magnetic field strengths while the jets' Lorentz factor is kept constant at 10. We use four different magnetic field strengths, $\beta=\infty$, 1, 1/2 and 1/4. Their corresponding magnetizations, $\sigma$ and their ratios between the Alfven speeds to the sound speeds, $M_a$ are summarized in the Table 1.

Fig. 10 shows the dispersion relation of the pinch mode ($m=0$) instabilities of the jets with various magnetizations. The red, green, blue and magenta lines are for the jets with mild ($\beta=\infty$), moderate ($\beta=1$), strong ($\beta=1/2$) and very strong ($\beta=1/4$) magnetization, respectively. Fig. 10(a) shows the evolution of the fundamental pinch mode instability. As the magnetic field strength becomes stronger, the fundamental pinch mode is stabilized at short as well as long wavelength. Fig. 10(a) shows that the fundamental pinch instability is already very stable for moderately relativistic jets ($\gamma_0=10$) without a magnetic field. The presence of magnetic field makes the jet even more stable at shorter and longer wavelengths. This is one of the highlights of the role of the magnetic field. Fig. 10(b) shows the evolution of the first reflection pinch mode instability. The magnetic field does not play a significant role in the stabilization of the first reflection pinch mode contrary to the fundamental pinch mode.

Fig. 11 is analogous to the Fig. 10 but for the kink mode ($m=1$) instabilities. Fig. 11(a) shows the evolution of the fundamental kink mode instability. In Fig. 11(a), the short wavelength fundamental kink instability modes are stabilized by the magnetic field. Stronger magnetization gives an increasingly improved stability at short wavelengths. The long wavelength modes, however, are not much affected by the magnetic field strength. On the other hand, the first reflection kink mode instability is stabilized at the short as well as the long wavelength by the strong magnetic field. Only very narrow range of wave numbers are above the stability criterion for the very strongly magnetized jet in Fig. 11(b).

There exists a cutoff in Figs. 10(a) and 11(a) at short wavelength for the strongly and very strongly magnetized jets. Let us define such a cutoff as a stability point where the temporal growth rate ($\omega_I$) is vanishing for both pinch and kink fundamental mode. In other words, the jets are stable when the wavelength of the perturbation is shorter than these stability points. This point corresponds to the onset of the bifurcation (or the mode splitting) in Kim *et al.* (2015). Notice that the real part of $\omega$ does not go to zero at the stability point. This fact is a distinguishable property from the marginal stability points of the reflection modes. (cf. both real and imaginary part of $\omega$ become zero at the marginal stability point of the reflection modes.) In fact, it does not mean that there is no fundamental mode at the wavelength longer than the stability point but there exists a stable fundamental mode whose $\omega_I$ is zero. Our numerical method is not capable of finding the stable modes. For that reason, we only show the unstable mode in Figs. 10(a) and 11(a). Interestingly, the fundamental mode becomes unstable again at short wavelength in the non-relativistic study (Kim *et al.* 2015). However, we cannot find such kind of unstable short wavelength fundamental mode branch for the strongly and very strongly magnetized jets in the wave number range that we study.

## 7 Discussion and Conclusions

Astrophysical jets such as AGN, GRB or protostellar jets are known to be surprisingly stable. They are able to propagate a distance over 10 million times of their initial radii before disruption. It is the most remarkable fact of the astrophysical jets compared to the terrestrial jets which can extend for few tens of their radii. In this paper, we investigate the impact of the magnetic field as well as the relativistic effects on the stability of jets in order to explain the noticeable stability of observed jets.

Most of the previous studies on the stability of relativistic jet dealt with the force-free approximation for simplicity. Or they assume a simplified jet structure to solve the linearized equations analytically. In this work, we forego any simplifying approximations. We linearize full relativistic MHD equation and solve it for the jets with realistic magnetic field configuration. Our numerical approach is one of the better ways of treating the equations without simplification.

We thoroughly studied the effects of the relativistic motion in the stability of the jets and verified that the relativistic effect significantly improves the stability of the jets. In other words, the higher Lorentz factor jets have the lower overall temporal growth rate. We also showed that the scaling law that is the relation between the maximum growth rate of the unstable modes and the Lorentz factor of the jets. The maximum temporal growth rates of the fundamental pinch instability mode are inversely proportion to the square of the Lorentz factor. However, the maximum temporal growth rates of all the other unstable modes are inversely proportion to the Lorentz factors. We presented the Lorentz factor up to 5000 only for the mathematical interests. However, the astrophysical jets usually have smaller Lorentz factors. For example, the Lorentz factor is ~50 for AGN jets, 100~1000 for GRB jets.

As shown in Table 2, the relativistic (proper) Mach number is nearly proportional to the Lorentz factor for the high Lorentz factor jets. Therefore, the Lorentz factor dependence on the growth rate of the jets can be reinterpreted as the growth rate as a function of proper Mach number. This relationship is also observed in the non-relativistic work i.e., compare Figs. 13 and 14 with Figs. 17 and 18 in Kim et al. (2015). We further obtain the exact relationship between the traditional Mach number and the temporal growth rate in the non-relativistic stability analysis and it is shown in Appendix D. We find the exactly same dependence as the relativistic result.

We also have investigated the stabilization by the realistic magnetic field, proposed by Gourgouliatos et al. (2012), that does not have current sheet on the surface of the jet. We have shown that the magnetic field in the jet also helps to improve the stability of the jet on top of the stabilizing effect by the relativistic speed. Highly magnetized jets have wide ranges of stability where the unstable modes are absent, which is consistent with the non-relativistic work by Kim *et al*. (2015).

In the dispersion relations of (non-)magnetized moderately relativistic ($\gamma_0$=10) jets, the maximum growth rates are around $10^{-1}$ except for the fundamental pinch mode which is negligible for the high Lorentz factor jets. The scaling relation suggests that the ultra-relativistic GRB jet with Lorentz factor of 1000 can propagate at least $4000r_j$ before undergoing one e-folding growth of the perturbation at the resonance wavelength. The AGN jet with Lorentz factor of 50, however, can propagate only $80r_j$. Although the relativistic effect as well as the magnetic field significantly improve the stability, it is still not enough to explain the observation (Piran 2004; Lister *et al*. 2009; Rykoff *et al*. 2009). One possibility to stabilize the jet is the lateral expansion of the jet. The lateral expansion can slow down the growth of the instability mode because it takes more time for the perturbation on the surface to propagate inside of jet if there is a lateral expansion (Rosen & Hardee 2000; Moll *et al*. 2008; Porth & Komissarov 2015).

As we discussed in our previous non-relativistic works (Kim et al. 2015, 2016), both Kelvin-Helmholtz (KH) and Current-Driven (CD) instabilities appear in all our results. To understand CD instabilities it is important to find the resonant surface where the resonance condition ($k_B=kB_{z0}+(m/r)B_{z0}=0$) holds. When this surface resides inside the jet, the jet becomes unstable to CD instability. Our model always has a resonant surface inside the jet regardless of $k$ and $m$ value. Therefore, it is not possible to isolate one contribution of instability from the other in the magnetized jet model that we study here. As a result, the non-magnetized or mildly magnetized (kinetically dominated) jets only show the KH instabilities, whereas both KH and CD instabilities appear in the jets with high magnetization.

In this paper, we only consider the jets with the top hat velocity profile. The jet with velocity shear inside has much improved stability in the non-relativistic linear stability analysis (Kim *et al*. 2016). We are planning to incorporate the velocity shear in the near future.

## Acknowledgements

DSB acknowledges support via NSF grants NSF-DMS-1361197, NSF-ACI-1533850, NSF-DMS-1622457. Several simulations were performed on a cluster at UND that is run by the Center for Research Computing. Computer support on NSF's XSEDE and Blue Waters computing resources is also acknowledged.

# APPENDIX A: Components of the Matrix A and B

In this appendix, the matrix components for solving our seven algebraic equations are shown. If we write the solution column vector as $\mathbf{X} = (\delta\rho, \delta P, \delta v_\phi, \delta v_z, \delta B_r, \delta B_\phi, \delta B_z)^\mathrm{T}$, our algebraic equations can be written in matrix form $\mathbf{AX=B}$. Then all the components of the matrix $\mathbf{A}$ are given as

$$\mathbf{A} = \begin{bmatrix} & & & \mathbf{A}_1 & & & \\ & & & \mathbf{A}_2 & & & \\ 0 & 0 & 0 & 0 & \tilde{\omega} & 0 & 0 \\ -i\tilde{\omega}\dfrac{B_{\phi 0}}{\rho_0} & 0 & ik_B & -i\tilde{\omega}\gamma_0^2 B_{\phi 0} v_{z0} & 0 & \tilde{\omega} & 0 \\ i\dfrac{k_B}{\rho_0} & 0 & -i\dfrac{k_B}{\tilde{\omega}}\dfrac{m}{r} & -i\gamma_0^2 \dfrac{k_B}{\tilde{\omega}}(k-\omega v_{z0}) & \dfrac{1}{r}+\dfrac{k}{\tilde{\omega}} v_{z0}' & -i\dfrac{m}{r} & ik \\ -\dfrac{\Gamma}{\rho_0} & \dfrac{1}{P_0} & 0 & 0 & 0 & 0 & 0 \\ 0 & -1 & -B_{\phi 0} B_{z0} v_{z0} & B_{\phi 0}^2 v_{z0} & 0 & -\dfrac{B_{\phi 0}}{\gamma_0^2} & -B_{z0} \end{bmatrix}, \quad (A1)$$

where the primed variables are their derivatives with respect to $r$. The row matrices $\mathbf{A}_1$ and $\mathbf{A}_2$ are

$$\mathbf{A}_1 = \begin{bmatrix} i\dfrac{B_{\phi 0}}{\rho_0}\left(\tilde{\omega} B_{z0} v_{z0} - \dfrac{k_B}{\gamma_0^2}\right) \\ 0 \\ i\tilde{\omega}\left(\rho_0 h_0 \gamma_0^2 + B_{z0}^2\right) + ik_B B_{z0} v_{z0} + i\dfrac{m}{r}\dfrac{k_B}{\tilde{\omega}}\dfrac{B_{\phi 0}}{\gamma_0^2} \\ iB_{\phi 0}(k-\omega v_{z0})\left(\dfrac{k_B}{\tilde{\omega}} - \gamma_0^2 B_{z0} v_{z0}\right) - i\omega B_{\phi 0} B_{z0} - 2i\dfrac{m}{r} B_{\phi 0}^2 v_{z0} \\ 2B_{\phi 0} v_{z0} v_{z0}' - \dfrac{1}{\gamma_0^2}\left(\dfrac{2}{r} B_{\phi 0} + B_{\phi 0}' + \dfrac{k}{\tilde{\omega}} B_{z0} v_{z0}'\right) \\ iB_{z0}(k-\omega v_{z0}) + 2i\dfrac{m}{r}\dfrac{B_{\phi 0}}{\gamma_0^2} \\ iB_{\phi 0}(k-\omega v_{z0}) \end{bmatrix},$$

$$\mathbf{A}_2 = \begin{bmatrix}
i\tilde{\omega}\gamma_0^2 v_{z0} - i\tilde{\omega}\dfrac{v_{z0}}{\rho_0}\left(\rho_0 h_0 \gamma_0^2 + B_{\phi 0}^2\right) - ik_B \dfrac{B_{z0}}{\rho_0} \\[6pt]
i\tilde{\omega}\gamma_0^2 v_{z0} \dfrac{\Gamma}{\Gamma-1} \\[6pt]
-i\tilde{\omega}B_{\phi 0}B_{z0} + ik_B\left(B_{\phi 0}v_{z0} + \dfrac{1}{\tilde{\omega}}\dfrac{m}{r}B_{z0}\right) \\[6pt]
i\tilde{\omega}\left(\rho_0 h_0 \gamma_0^4 + B_{z0}^2\right) + i\gamma_0^2(k - \omega v_{z0})\left(\dfrac{k_B}{\tilde{\omega}}B_{z0} - B_{\phi 0}^2 v_{z0}\right) \\[6pt]
-\dfrac{B_{z0}}{r} - B'_{z0} - \dfrac{k}{\tilde{\omega}}B_{z0}v'_{z0} \\[6pt]
2i\tilde{\omega}B_{\phi 0}v_{z0} + i\dfrac{m}{r}B_{z0} \\[6pt]
i(k_B + kB_{z0})
\end{bmatrix}^{\mathrm{T}}.$$

The components of column vector $\mathbf{B}$ are

$$\mathbf{B} = \begin{bmatrix}
\left[\dfrac{1}{r^2}\left(r^2 B_{\phi 0}B_{z0}v_{z0}\right)' + \dfrac{B_{\phi 0}}{\gamma_0^2 \tilde{\omega}}\left(k_B \dfrac{(r\rho_0\gamma_0)'}{r\rho_0\gamma_0} - k'_B\right) - B_{\phi 0}B_{z0}v_{z0}\dfrac{(r\rho_0\gamma_0)'}{r\rho_0\gamma_0}\right]\delta v_r + i\dfrac{m}{r}\delta\Pi \\[6pt]
\left[-\dfrac{1}{r}\left[r\left(\rho_0 h_0 \gamma_0^2 + B_{\phi 0}^2\right)v_{z0}\right]' + \dfrac{B_{z0}}{\tilde{\omega}}\left(k_B \dfrac{(r\rho_0\gamma_0)'}{r\rho_0\gamma_0} - k'_B\right) + \left(\rho_0 h_0 \gamma_0^2 + B_{\phi 0}^2\right)v_{z0}\dfrac{(r\rho_0\gamma_0)'}{r\rho_0\gamma_0}\right]\delta v_r + ik\delta\Pi \\[6pt]
-ik_B \delta v_r \\[6pt]
\left(B_{\phi 0}\dfrac{(r\rho_0\gamma_0)'}{r\rho_0\gamma_0} - B'_{\phi 0}\right)\delta v_r \\[6pt]
-\dfrac{1}{\tilde{\omega}}\left(k_B \dfrac{(r\rho_0\gamma_0)'}{r\rho_0\gamma_0} - k'_B\right)\delta v_r \\[6pt]
0 \\[6pt]
-\delta\Pi
\end{bmatrix}$$

. (A2)

Note that the components of matrix $\mathbf{A}$ consist only of the unperturbed variables and the components of matrix $\mathbf{B}$ have $\delta v_r$, $\delta\Pi$ in addition to the unperturbed variables. Once we know the values of $\delta v_r$, $\delta\Pi$, which are the solutions of differential equations (8) and (9), we can obtain all the other perturbations that is explicitly written in $\mathbf{X}$ by solving the matrix equation $\mathbf{AX}=\mathbf{B}$.

# APPENDIX B: ASYMPTOTIC BEHAVIOR OF SOLUTIONS AT SMALL RADII

In order to obtain a solution $\delta v_r$ and $\delta\Pi$ of the differential equations in (8) and (9), we should know the asymptotic behavior of $\delta v_r$ and $\delta\Pi$ near the axis. Then we can start integration from $r=0$. The power laws of all the perturbation variables deduced from linearized equation for the pinch mode ($m = 0$) by assuming $\delta\rho \sim r^\alpha$ near $r = 0$ are provided as:

$$\delta\rho = \delta\rho^* r^\alpha,\ \delta P = \delta P^* r^\alpha,\ \delta\Pi = \delta\Pi^* r^\alpha,$$
$$\delta v_r = \delta v_r^* r^{\alpha+1},\ \delta v_\phi = \delta v_\phi^* r^{\alpha+1},\ \delta v_z = \delta v_z^* r^\alpha,\ \delta B_r = \delta B_r^* r^{\alpha+1},\ \delta B_\phi = \delta B_\phi^* r^{\alpha+1},\ \delta B_z = \delta B_z^* r^\alpha. \tag{B1}$$

Up to leading order (after cancelling out leading order of $r$) Eqs. (8)-(16) become

$$i\tilde\omega \frac{\delta\rho^*}{\rho_0} + (\alpha+2)\delta v_r^* + i(\tilde\omega\gamma_0 v_{z0} - k)\delta v_z^* = 0, \tag{B2}$$

$$\alpha\delta\Pi^* = 0, \tag{B3}$$

$$i\tilde\omega\rho_0 h_0 \gamma_0^2 \delta v_\phi^* + i\omega\left(B_{z0}^2 \delta v_\phi^* - B'_{\phi 0}B_{z0}\delta v_z^* - B_{z0}v_{z0}\delta B_\phi^* - B'_{\phi 0}v_{z0}\delta B_z^*\right)$$
$$-(\alpha-4)B'_{\phi 0}\left[B_{z0}v_{z0}\delta v_r^* - (1-v_{z0}^2)\delta B_\phi^*\right] + ik\left(B'_{\phi 0}\delta B_z^* + B_{z0}\delta B_\phi^*\right) = 0, \tag{B4}$$

$$i\tilde\omega\left[\rho_0 h_0 \gamma_0^2 (1+2\gamma_0^2 v_{z0}^2)\delta v_z^* + \gamma_0^2 v_{z0}\left(\delta\rho^* + \frac{\Gamma}{\Gamma-1}\delta P^*\right)\right]$$
$$+(\alpha+2)\left(\rho_0 h_0 \gamma_0^2 v_{z0}\delta v_r^* - B_{z0}\delta B_r^*\right) - ik\rho_0 h_0 \gamma_0^2 v_{z0}\delta v_z^* + 2ikB_{z0}\delta B_z^* - ik\delta\Pi^* = 0 \tag{B5}$$

$$\tilde\omega\delta B_r^* + kB_{z0}\delta v_r^* = 0, \tag{B6}$$

$$i\tilde\omega\delta B_\phi^* + (\alpha+2)B'_{\phi 0}\delta v_r^* - ik\left(B'_{\phi 0}\delta v_z^* - B_{z0}\delta v_\phi^*\right) = 0, \tag{B7}$$

$$(\alpha+2)\delta B_r^* - ik\delta B_z^* = 0, \tag{B8}$$

$$\rho_0\delta P^* - \Gamma P_0 \delta\rho^* = 0, \tag{B9}$$

$$\delta\Pi^* = \delta P^* + B_{z0}\delta B_z^*. \tag{B10}$$

Since $\delta\Pi^*$ can be expressed in terms of $v_r^*$, $\alpha$ must be zero for the non-trivial solution of $\delta v_r^*$. To find the expression of $\delta\Pi^*$ in Eq. (B10), we need to know the expression of $\delta P^*$ and $\delta B_z^*$ in terms of $\delta v_r^*$. However, there is no simple way to get $\delta P^*$ because Eqs. (B2)-(B9) are completely coupled each other. After lengthy manipulation, we can find the expression of $\delta\Pi^*$:

$$\frac{\delta\Pi^*}{\delta v_r^*} = \frac{2iB_{z0}^2}{\tilde{\omega}} - \frac{2i\tilde{\omega}\rho_0 h_0 \gamma_0^2}{\tilde{\omega}\gamma_0^2\left(1-1/c_s^2\right)-\left(\omega^2-k^2\right)}. \tag{B11}$$

Like *m*=0, pinch mode, the perturbation variables have following relations near $r=0$ for the kink mode (*m*=1):

$$\delta\rho = \delta\rho^* r^\alpha, \ \delta P = \delta P^* r^\alpha, \ \delta\Pi = \delta\Pi^* r^\alpha,$$
$$\delta v_r = \delta v_r^* r^{\alpha-1}, \ \delta v_\phi = \delta v_\phi^* r^{\alpha-1}, \ \delta v_z = \delta v_z^* r^\alpha, \ \delta B_r = \delta B_r^* r^{\alpha-1}, \ \delta B_\phi = \delta B_\phi^* r^{\alpha-1}, \ \delta B_z = \delta B_z^* r^\alpha. \tag{B12}$$

Substitution of above relations in Eqs. (8)-(16) gives

$$\alpha\delta v_r^* - im\delta v_\phi^* = 0, \tag{B13}$$

$$\alpha\delta\Pi^* + i\tilde{\omega}\rho_0 h_0 \gamma_0^2 \delta v_r^* + ik_B^* \delta B_r^* + i\left(\omega B_{z0} + mB'_{\phi 0} v_{z0}\right)\left(\omega B_{z0}\delta v_r^* + v_{z0}\delta B_r^*\right)$$
$$+2B'_{\phi 0}\left[\left(1-v_{z0}^2\right)B_\phi^* + B_{z0} v_{z0}\delta v_\phi^*\right] = 0 \tag{B14}$$

$$i\tilde{\omega}\rho_0 h_0 \gamma_0^2 \delta v_\phi^* + i\omega\left(B_{z0}^2 \delta v_\phi^* - B_{z0}v_{z0}\delta B_\phi^*\right) + 2imB'_{\phi 0}\left[B_{z0}v_{z0}\delta v_\phi^* + \left(1-v_{z0}^2\right)\delta B_\phi^*\right] - im\delta\Pi^*$$
$$+ikB_{z0}\delta B_\phi^* - (\alpha+2)B'_{\phi 0}\left[B_{z0}v_{z0}\delta v_r^* + \left(1-v_{z0}^2\right)\delta B_r^*\right] = 0 \tag{B15}$$

$$\alpha\left(\rho_0 h_0 \gamma_0^2 v_{z0}\delta v_r^* - B_{z0}\delta B_r^*\right) - im\left(\rho_0 h_0 \gamma_0^2 v_{z0}\delta v_\phi^* - B_{z0}\delta B_\phi^*\right) = 0, \tag{B16}$$

$$\tilde{\omega}\delta B_r^* + k_B^* \delta v_r^* = 0, \tag{B17}$$

$$i\tilde{\omega}\delta B_\phi^* + \alpha B'_{\phi 0}\delta v_r^* + ikB_{z0}\delta v_\phi^* = 0, \tag{B18}$$

$$\alpha\delta B_r^* - im\delta B_\phi^* = 0, \tag{B19}$$

$$\rho_0 \delta P^* - \Gamma P_0 \delta\rho^* = 0, \tag{B20}$$

$$\delta\Pi^* = \delta P^* + B_{\phi 0}\delta B_\phi^* + B_{z0}\delta B_z^* - B_{\phi 0} v_{z0}\left(v_{z0}\delta B_\phi^* - B_{z0}\delta v_\phi^*\right), \tag{B21}$$

where $k_B^* = mB'_{\phi 0} + kB_{z0}$. Eqs. (B13), (B17), (B18) and (B19) give the expressions of $\delta v_\phi^*$, $\delta B_r^*$ and $\delta B_\phi^*$ in terms of $\delta v_r^*$. Then, we substitute the expression of $\delta v_\phi^*$, $\delta B_r^*$ and $\delta B_\phi^*$ in Eq. (B14) and obtain the expression of the total pressure perturbation ($\delta\Pi^*$) in terms of $\delta v_r^*$. Finally, we can make the angular momentum equation (B15) dependent only on $\delta v_r^*$:

$$\frac{\alpha^2 - m^2}{\alpha m}\left[\tilde{\omega}\rho_0 h_0 \gamma_0^2 + \omega B_{z0}\left(B_{z0} + v_{z0}\frac{k_B^*}{\tilde{\omega}}\right) + mv_{z0}\left(B_{\phi 0}'B_{z0} + \frac{k_B^*}{\tilde{\omega}}\right) - \frac{\left(k_B^*\right)^2}{\tilde{\omega}}\right]\delta v_r^* = 0. \qquad (B22)$$

In Eq. (B22), $\delta v_r^*$ has non-trivial solution when $\alpha = \pm m$. Accordingly, we only take $\alpha = |m|$ solution which is not diverging at $r = 0$. Then, we can find the final expression of $\delta \Pi^*$:

$$\frac{\delta \Pi^*}{\delta v_r^*} = -\frac{i}{|m|}\left[\tilde{\omega}\rho_0 h_0 \gamma_0^2 - \frac{\left(k_B^*\right)^2}{\tilde{\omega}} + \left(\omega B_{z0} + mB_{\phi 0}'v_{z0}\right)\left(B_{z0} + v_{z0}\frac{k_B^*}{\tilde{\omega}}\right)\right] - \frac{2iB_{\phi 0}'}{m}\left[\left(1 - v_{z0}^2\right)\frac{k_B^*}{\tilde{\omega}} - B_{z0}v_{z0}\right]. \qquad (B23)$$

**Appendix C: Dispersion relation and Marginal Stability points of the non-magnetized jets**

The non-magnetized jet has uniform density, pressure. If the velocity profile is top-hat, we can obtain an analytic interior solution which is also expressed as the modified Bessel function $I_{|m|}$ just like the outside solution. Then the boundary condition gives a following equation:

$$\frac{\tilde{\omega}_j^2 \rho_j h_j \gamma_j^2 I_{|m|}(\kappa_j)}{\kappa_j I_{|m|}'(\kappa_j)} = \frac{\omega^2 \rho_a h_a K_{|m|}(\kappa_a)}{\kappa_a K_{|m|}'(\kappa_a)}, \qquad (C1)$$

where $\kappa_j = \gamma_j\sqrt{\left(k - \omega v_j\right)^2 - \tilde{\omega}_j^2/c_{s,j}^2}$ and $\kappa_a = \sqrt{k^2 - \omega^2/c_{s,a}^2}$. The subscripts $j$ and $a$ denote the jet and ambient medium, respectively.

The marginal stability points of the reflection modes are the $k$ value where the real and imaginary part of $\omega$ becomes zero (Cohn 1983). Then the Eq. (C1) reduces to

$$J_{|m|}\left(\gamma_j k \sqrt{M_j^2 - 1}\right) = 0, \qquad (C2)$$

where $M_j$ is the Mach number of the jet. We keep jets' Mach number constant at 4. Therefore, the wave number $k_n$ at the marginal stability point for the n-th reflection mode is inversely proportional to the Lorentz factor ($k_n \sim 1/\gamma_j$).

**Appendix D: Scaling Relation in the Non-Relativistic Stability Analysis**

In this appendix D, we present the non-relativistic counterparts of the scaling relationships that are shown in section 4.3 and 5.2. The strongly relativistic jet ($\gamma_0$>>1) has a proper Mach number which is proportional to its Lorentz factor (see Eq. (18)). The definition of the Mach number used in this paper for the Mach number differs from the definition of the proper Mach number for a relativistic jet. If we had used the proper Mach number for the relativistic jets, the scaling relation between the maximum temporal growth rate and the Lorentz factor is indeed same as the scaling relation between the maximum temporal growth rate and the proper Mach number. It is important to see that there exist corresponding relationships for the non-relativistic jets i.e., the growth rate as a function of Mach number. Fig. A1 shows the scaling relations of the non-magnetized jets. The maximum temporal growth rates of the fundamental pinch instability mode are inversely proportion to the square of the Mach number. However, the maximum temporal growth rates of all the other unstable modes are inversely proportion to the Mach numbers. It follows exactly same scaling relation as in Fig. 5.

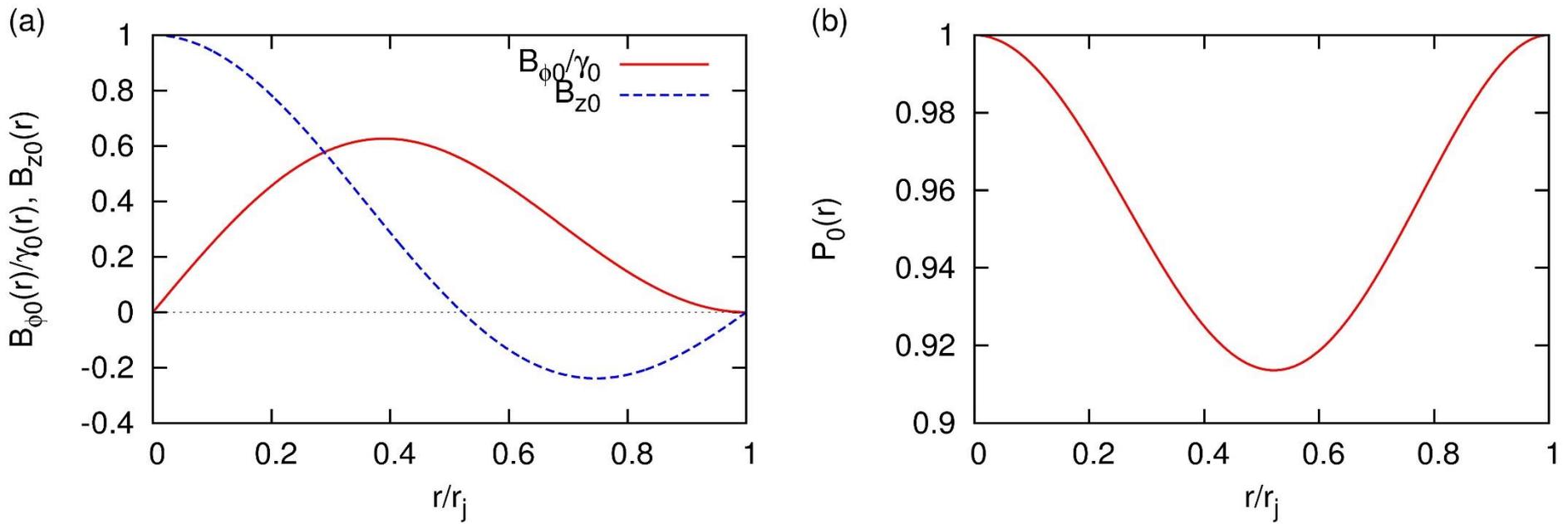

*Figure 1. (a) The toroidal magnetic field (red solid line) and the axial field (blue dashed line) as a function of the jet radius from Gourgouliatos et al. (2012). Notice that the toroidal magnetic field is proportional to the Lorentz factor of the jet. The fields are zero at the jet boundary, resulting in jets that do not have a current sheet at the boundary. (b) The corresponding gas pressure in the jet as a function of jet radius.*

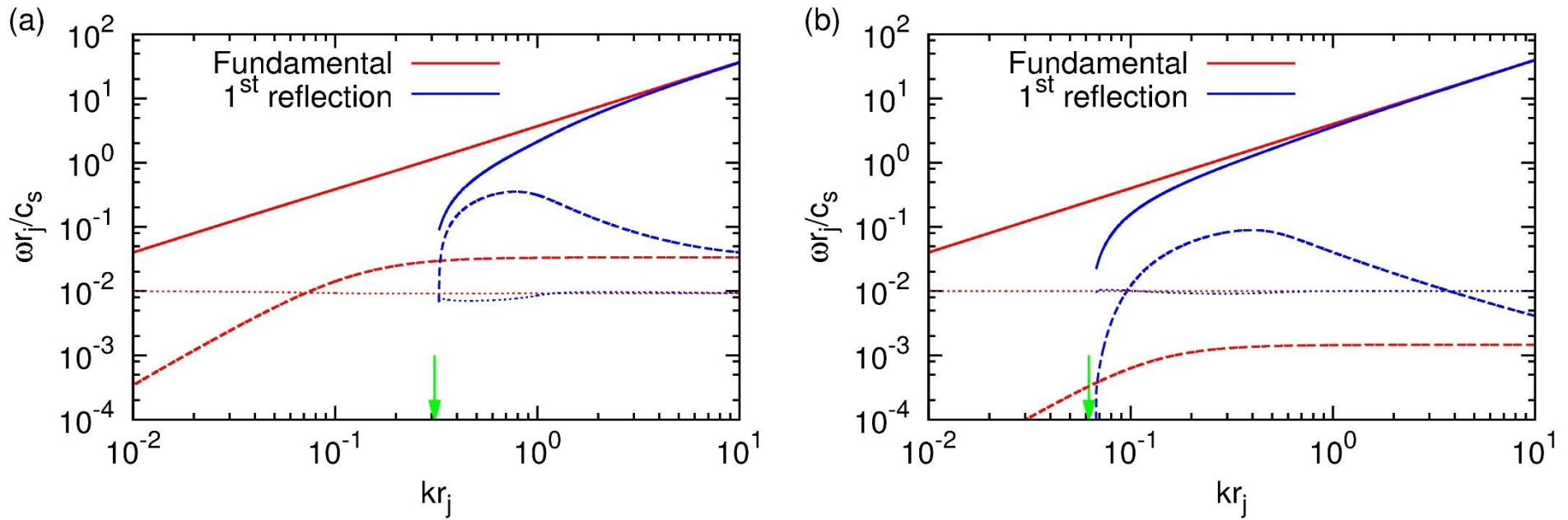

Figure 2. The dispersion relation of the pinch mode (m=0) instabilities for the non-magnetized relativistic jets. (a) The unperturbed jet has mild ($\gamma_0=2$) Lorentz factor. (b) The corresponding figure for the jet with moderate ($\gamma_0=10$) Lorentz factor. The red and blue lines represent the fundamental and the first reflection mode. The real and imaginary part of $\omega$ are shown as the solid and dashed lines. The dotted curves refer to the stability criterion for each of the unstable mode. The green arrows indicate the marginal stability point of the first reflection modes.

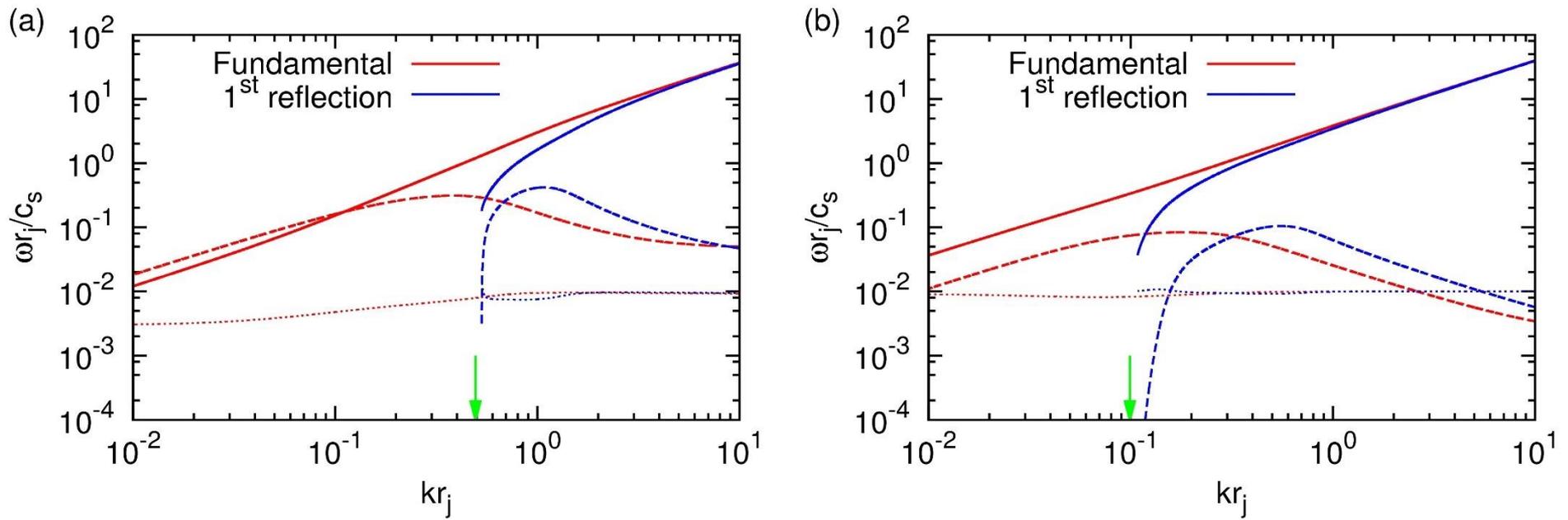

*Figure 3. The dispersion relation of the kink mode (m=1) instabilities for the non-magnetized relativistic jets. (a) The unperturbed jet has mild ($\gamma_0=2$) Lorentz factor. (b) The corresponding figure for the jet with moderate ($\gamma_0=10$) Lorentz factor. The red and blue lines represent the fundamental and the first reflection mode. The real and imaginary part of $\omega$ are shown as the solid and dashed lines. The dotted curves refer to the stability criterion for each of the unstable mode. The green arrows indicate the marginal stability point of the first reflection modes.*

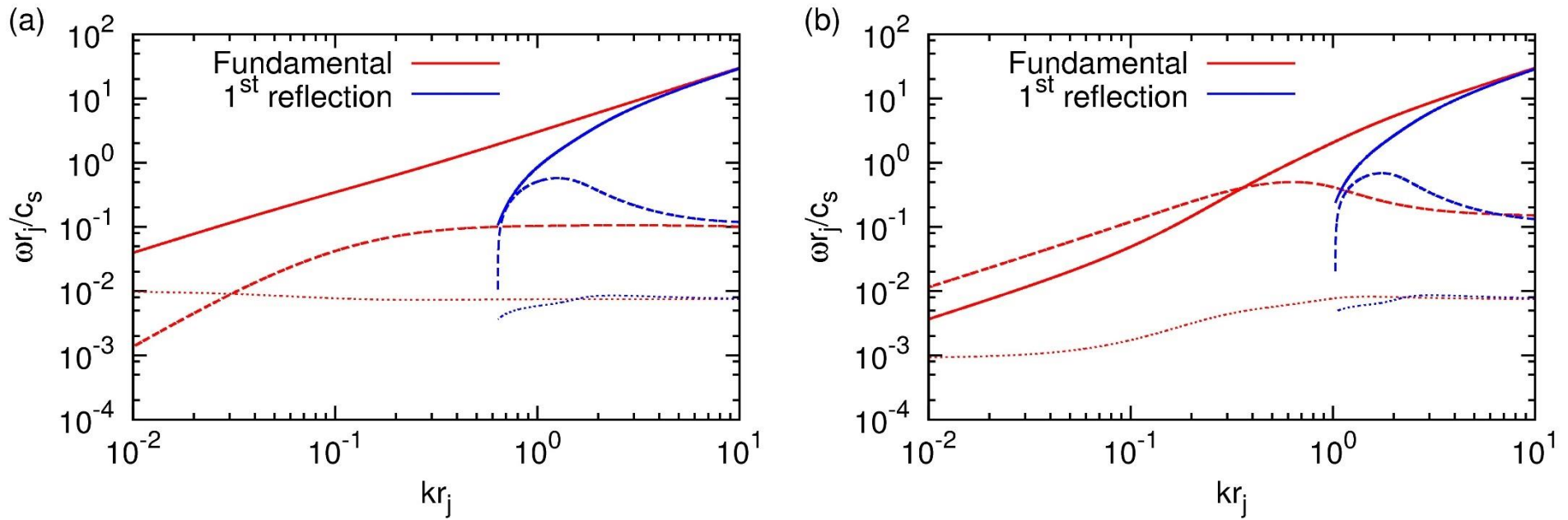

*Figure 4. The dispersion relation of the non-relativistic non-magnetized ($\beta=\infty$) jets with $\eta=0.1$ and $M=4$. The velocity profile in the jet is top-hat. This is obtained in the non-relativistic linear stability analysis which appears in Kim et al. (2015). (a) shows the pinch mode ($m=0$) instabilities while (b) shows the kink mode ($m=1$) instabilities. The red and blue lines represent the fundamental and the first reflection mode. The real and imaginary part of $\omega$ are shown as the solid and dashed lines. The dotted curves refer to the stability criterion for each of the unstable mode.*

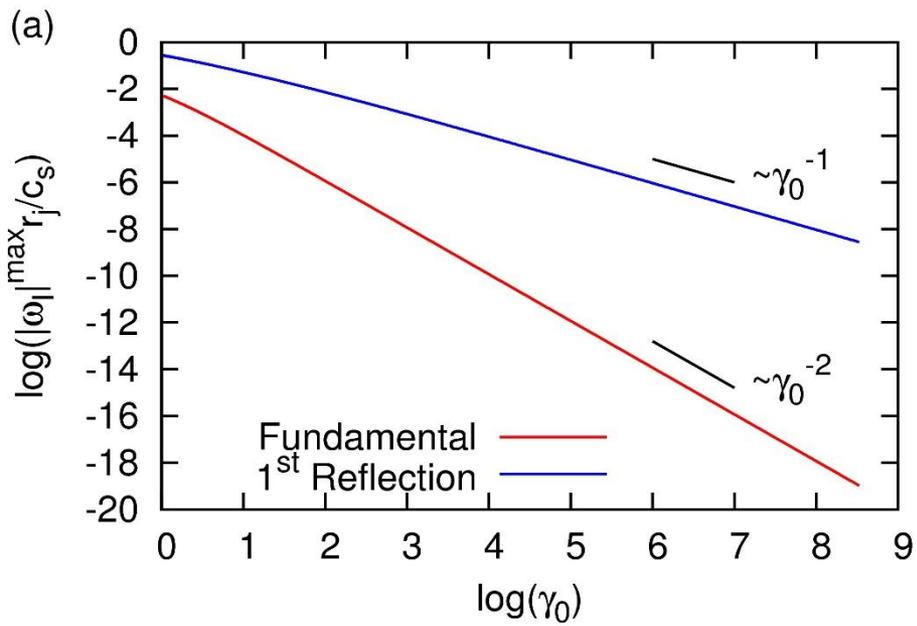 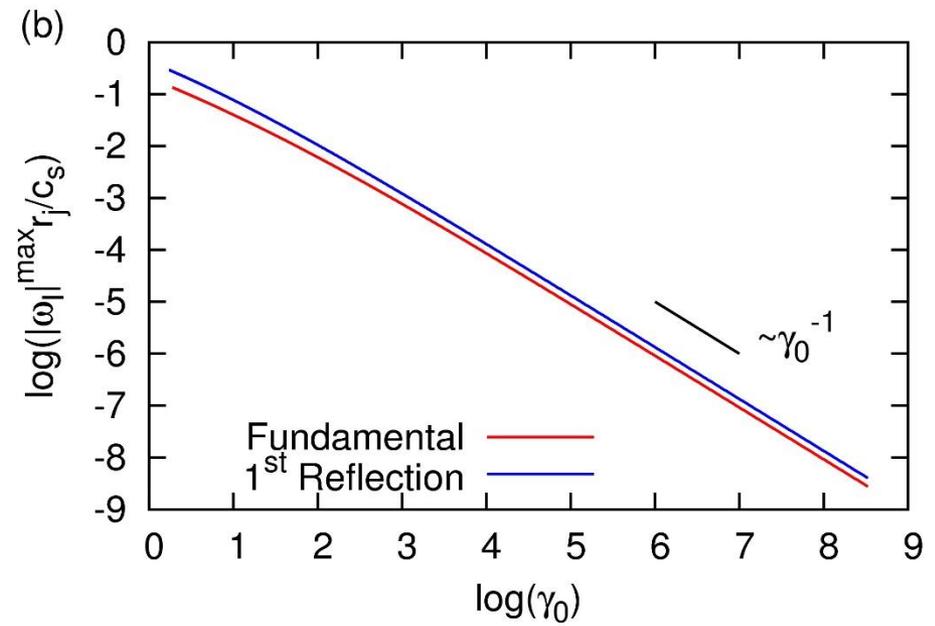

*Figure 5. The scaling property of the non-magnetized jets with increasing Lorentz factor. (a) shows the pinch mode (m=0) instabilities while (b) shows the kink mode (m=1) instabilities. The red and blue lines represent the fundamental and the first reflection mode.*

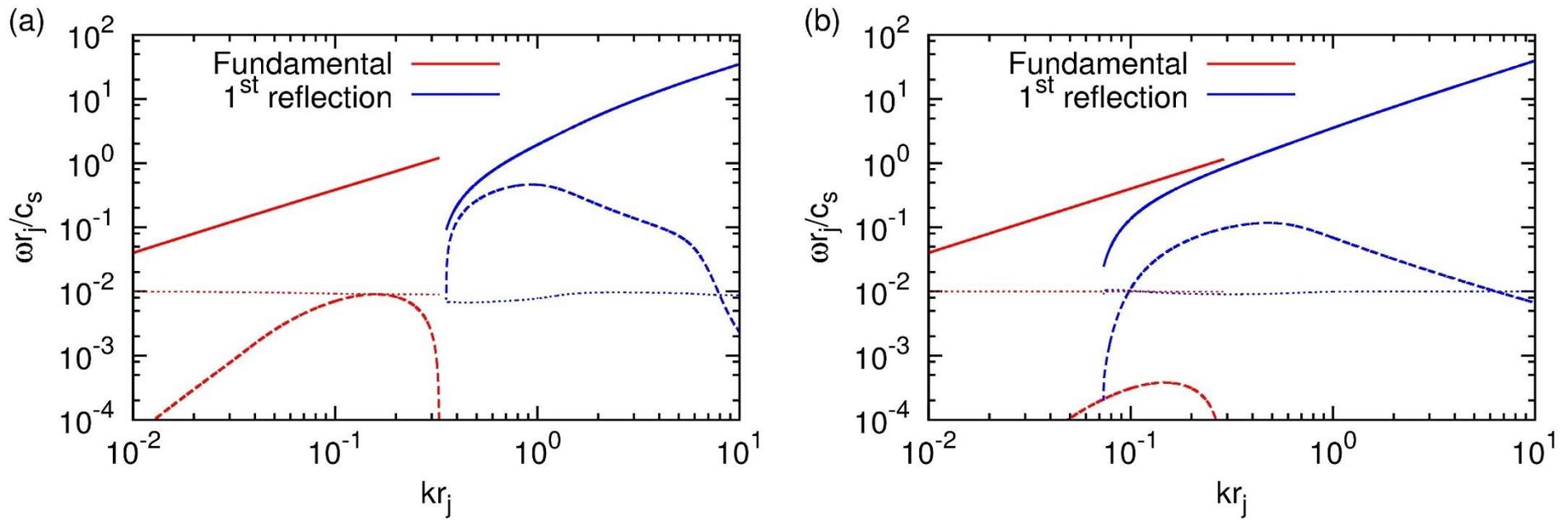

Figure 6. The dispersion relation of the pinch mode (m=0) instabilities for the magnetized relativistic jets. The plasma β of the jet is 1/2. (a) The unperturbed jet has mild ($\gamma_0=2$) Lorentz factor. (b) The corresponding figure for the jet with moderate ($\gamma_0=10$) Lorentz factor. The red and blue lines represent the fundamental and the first reflection mode. The real and imaginary part of ω are shown as the solid and dashed lines. The dotted curves refer to the stability criterion for each of the unstable mode.

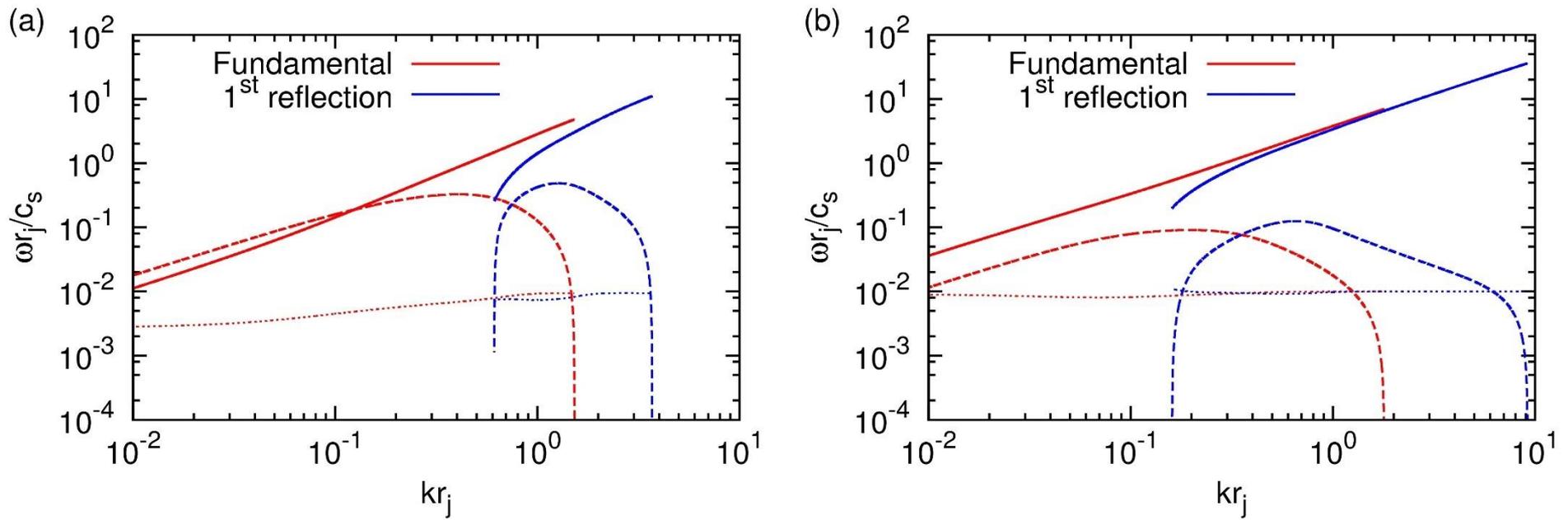

*Figure 7. The dispersion relation of the kink mode (m=1) instabilities for the magnetized relativistic jets. The plasma β of the jet is 1/2. (a) The unperturbed jet has mild ($\gamma_0=2$) Lorentz factor. (b) The corresponding figure for the jet with moderate ($\gamma_0=10$) Lorentz factor. The red and blue lines represent the fundamental and the first reflection mode. The real and imaginary part of ω are shown as the solid and dashed lines. The dotted curves refer to the stability criterion for each of the unstable mode.*

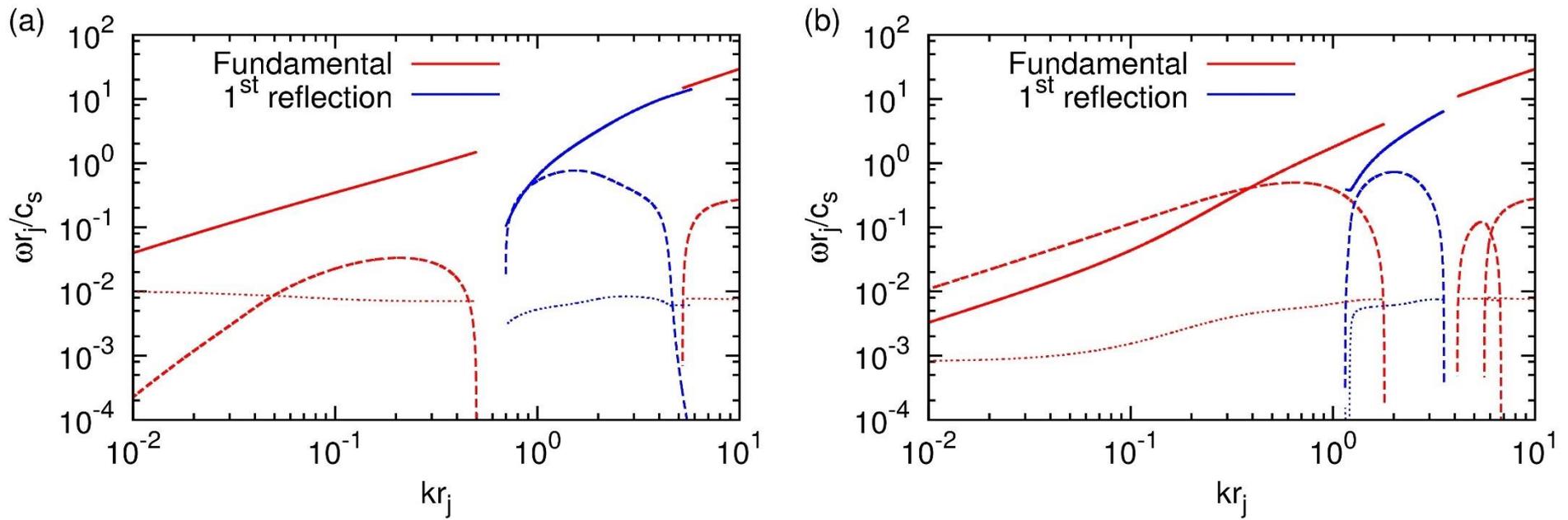

*Figure 8. The dispersion relation of the non-relativistic magnetized jets with η=0.1 and M=4. The velocity profile in the jet is top-hat. The plasma β of the jet is 1/2. This is obtained in the non-relativistic linear stability analysis which appears in Kim et al. (2015). (a) shows the pinch mode (m=0) instabilities while (b) shows the kink mode (m=1) instabilities. The red and blue lines represent the fundamental and the first reflection mode. The real and imaginary part of ω are shown as the solid and dashed lines. The dotted curves refer to the stability criterion for each of the unstable mode.*

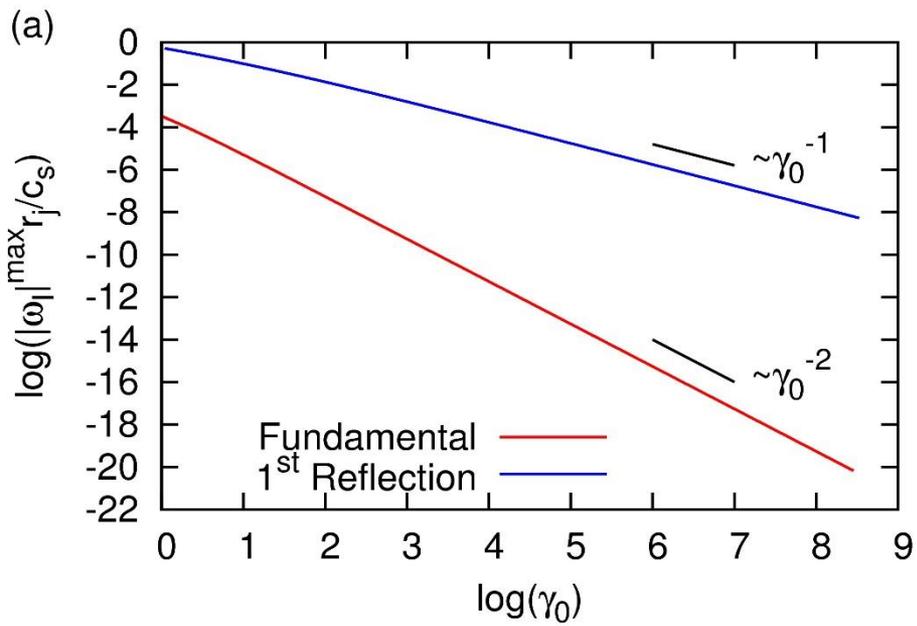 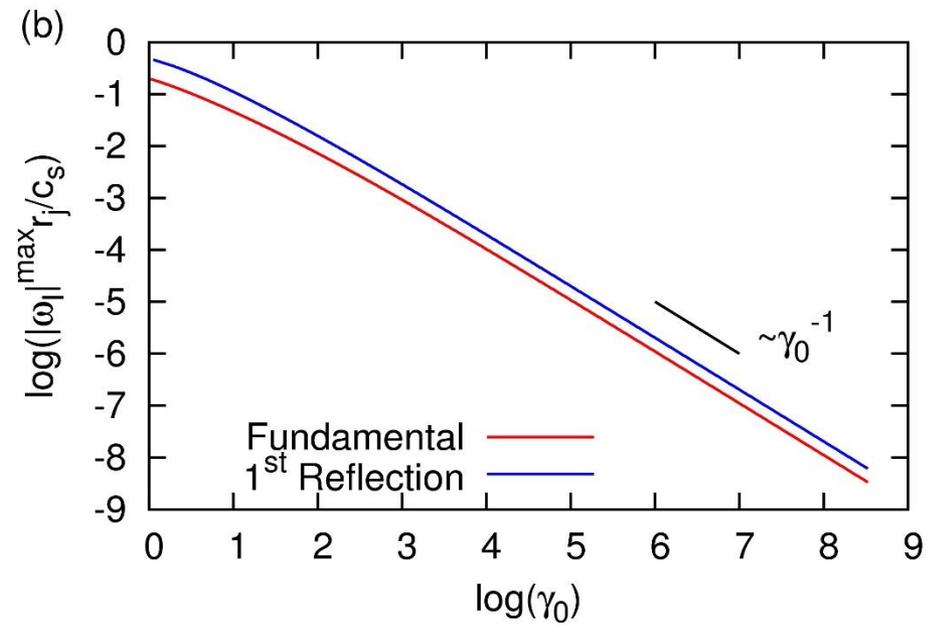

*Figure 9. The scaling property of the strongly magnetized jets (β=1/2) with increasing Lorentz factor. (a) shows the pinch mode (m=0) instabilities while (b) shows the kink mode (m=1) instabilities. The red and blue lines represent the fundamental and the first reflection mode.*

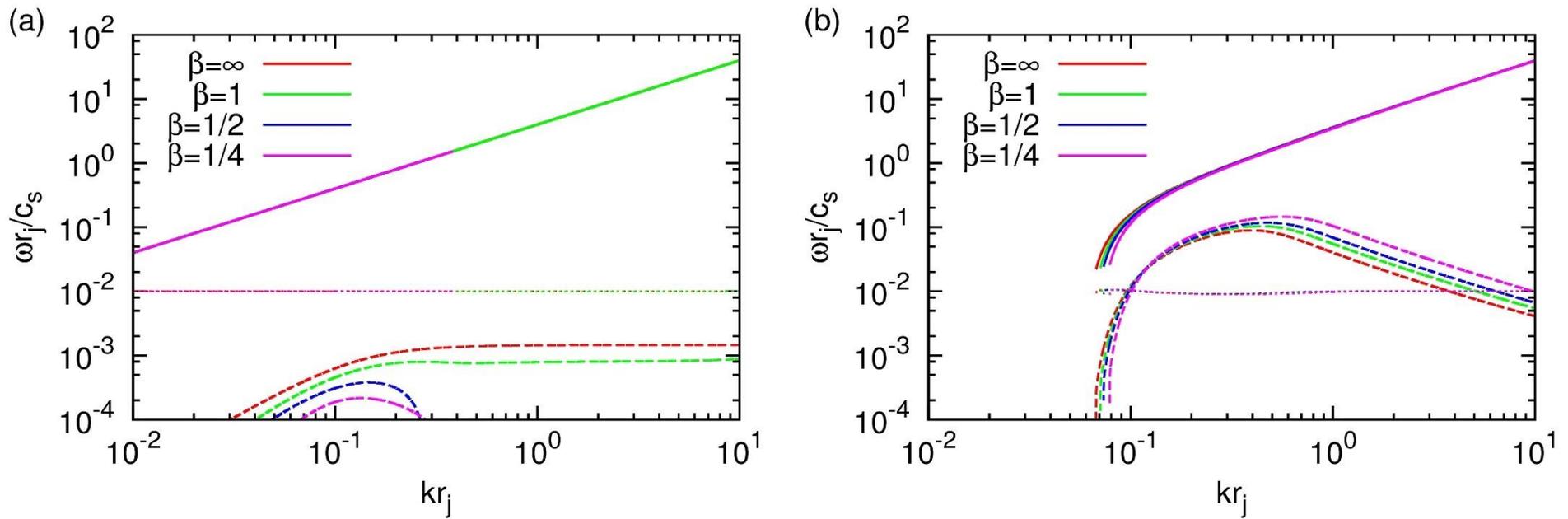

Figure 10. The dispersion relation of the pinch mode (m=0) instabilities for relativistic jets with various magnetizations. The unperturbed jet has the Lorentz factor of 10. The red, green, blue and magenta lines are for the jets with mild ($\beta=\infty$), moderate ($\beta=1$), strong ($\beta=1/2$) and very strong ($\beta=1/4$) magnetization. (a) shows the fundamental mode instability while (b) shows the first reflection mode instability.

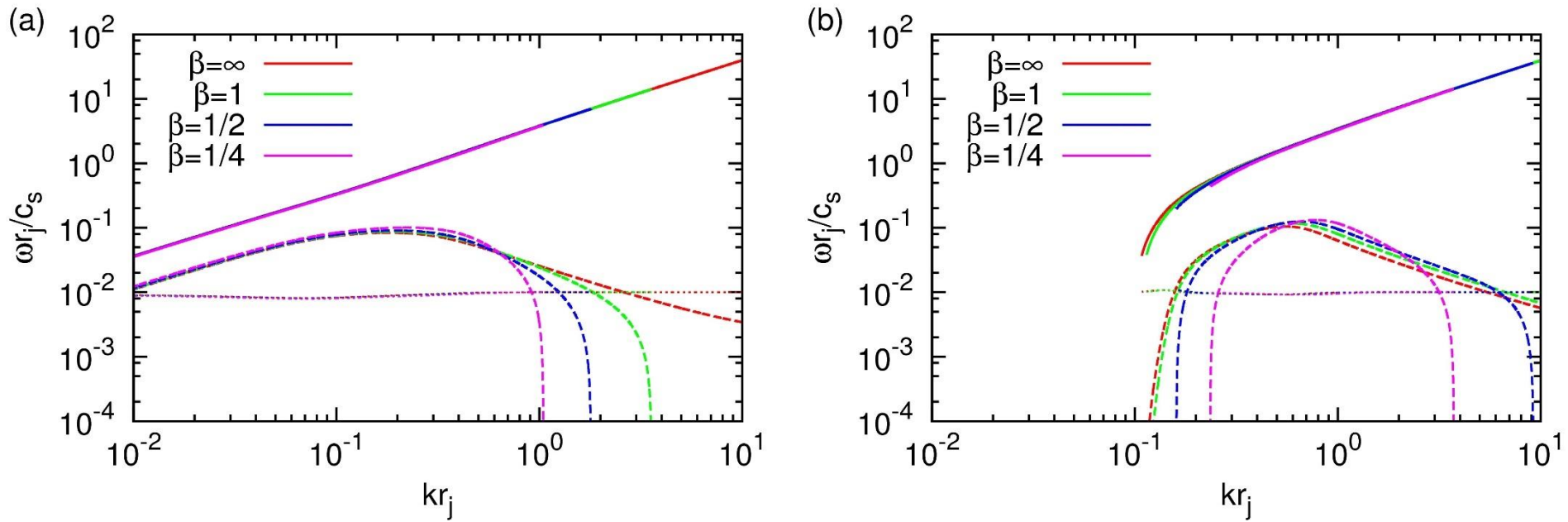

*Figure 11. The dispersion relation of the kink mode (m=1) instabilities for relativistic jets with various magnetizations. The unperturbed jet has the Lorentz factor of 10. The red, green, blue and magenta lines are for the jets with mild (β=∞), moderate (β=1), strong (β=1/2) and very strong (β=1/4) magnetization. (a) shows the fundamental mode instability while (b) shows the first reflection mode instability.*

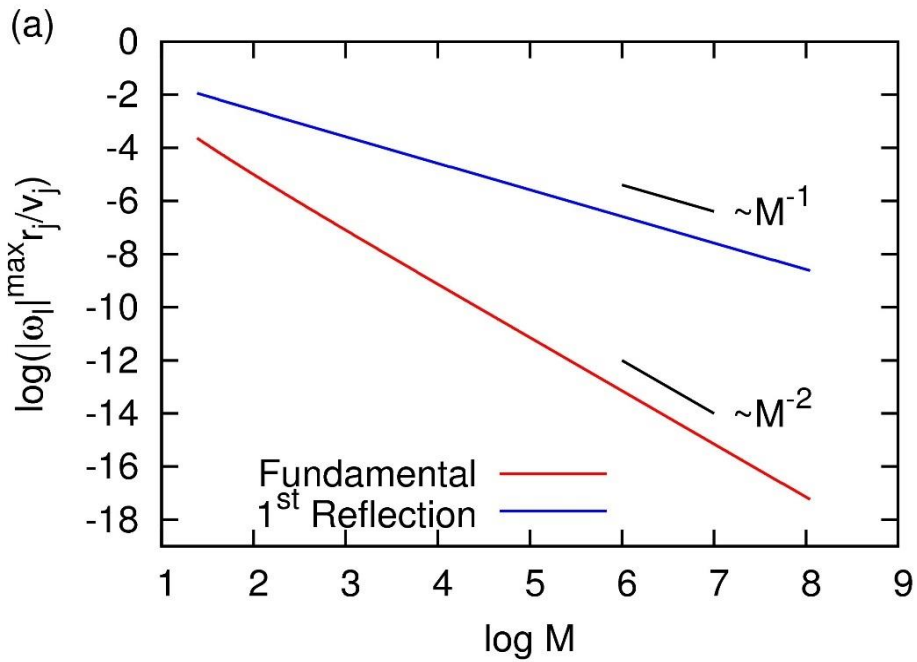 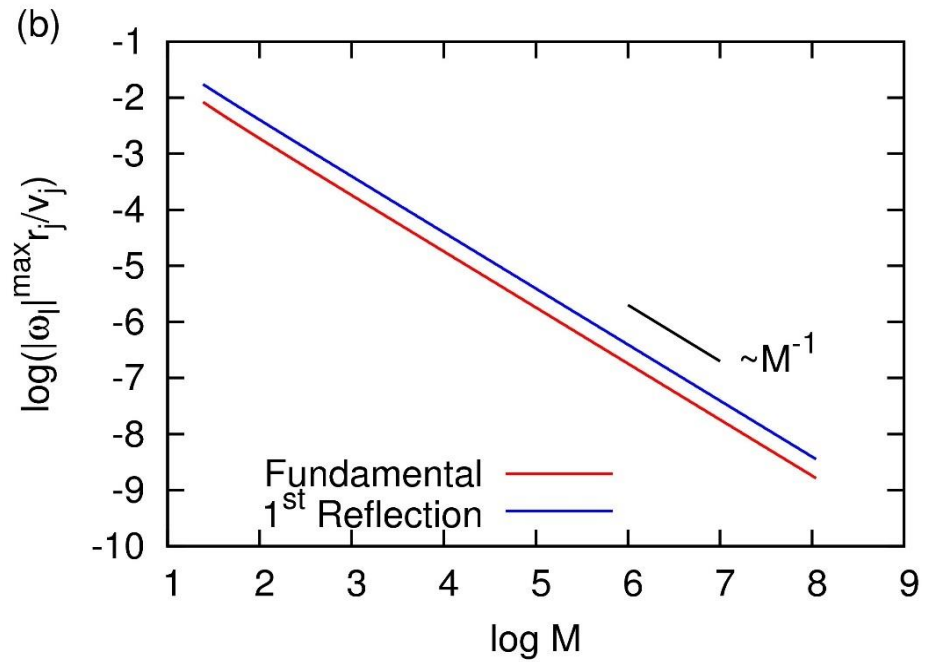

*Figure A1. The scaling property of the non-relativistic non-magnetized jets with increasing Mach number. (a) shows the pinch mode (m=0) instabilities while (b) shows the kink mode (m=1) instabilities. The red and blue lines represent the fundamental and the first reflection mode.*